\documentclass[preprint,12pt]{elsarticle}

\usepackage{listings}
\usepackage{graphicx}
\usepackage{booktabs} 
\usepackage{array,multirow}
\usepackage{url} 
\usepackage{natbib}
\usepackage{amssymb}
\usepackage{lscape}
\usepackage{mathrsfs}
\usepackage{subfig}
\usepackage{amsmath}
\usepackage{tikz}
\usepackage{float}
\usepackage{rotating}
\usepackage{ulem}
\usepackage{algorithm}
\usepackage[noend]{algpseudocode}

\usepackage{hyperref}
\usepackage{tikz}
\usetikzlibrary{arrows,shapes,positioning,shadows,trees}
\usepackage{todonotes}
\usepackage{soul}

\tikzset{
  basic/.style  = {draw, text width=4cm, drop shadow, font=\sffamily, rectangle},
  root/.style   = {basic, rounded corners=2pt, thin, align=center,fill=green!60},
  level 2/.style = {basic, rounded corners=6pt, thin,align=center, fill=green!30,text width=8em},
  level 3/.style = {basic, thin, align=left, fill=pink!60, text width=6.5em},
  level 3a/.style = {basic, thin, align=left, fill=pink!30, text width=6.5em},
  level 4/.style = {basic, thin, align=left, fill=gray!40, text width=6.5em},
  level 5/.style = {basic, thin, align=center, fill=gray!5,text width=6.5em}
}

\makeatletter
\providecommand{\doi}[1]{%
	\begingroup
	\let\bibinfo\@secondoftwo
	\urlstyle{rm}%
	\href{http://dx.doi.org/#1}{%
		doi:\discretionary{}{}{}%
		\nolinkurl{#1}%
	}%
	\endgroup
}
\makeatother

\usepackage{amssymb}

\journal{Journal of Network and Computer Applications} 

\begin{document}

\begin{frontmatter}






\title{Genetic-based fog colony optimization hybridized with hierarchical clustering and its influence in the placement of fog services}


\author{Francisco Talavera\corref{}}
\ead{f.talavera@uib.es}

\author{Isaac Lera\corref{}}
\ead{isaac.lera@uib.es}

\author{Carlos Juiz\corref{}}
\ead{cjuiz@uib.es}

\author{Carlos Guerrero\corref{mycorrespondingauthor}}
\ead{carlos.guerrero@uib.es}
\cortext[mycorrespondingauthor]{Corresponding author}

\address{Crta. Valldemossa km 7.5, Palma, E07121, SPAIN}

\address[mymainaddress]{Computer Science Department, University of Balearic Islands}

\begin{abstract}

The organization of fog devices into fog colonies has reduced the complexity management of fog domains. One of the main influencing factors on this complexity is the large number of devices, i.e. the high scale level of the infrastructure. Fog colonies are subsets of fog devices that are managed independently from the other colonies. Thus, the number of devices involved in the management of a colony is much smaller. Previous studies have evaluated the influence of the fog colony layout on system performance metrics. We propose to use a hierarchical clustering as the base definition of the fog colony layout of the fog infrastructure. The dendrogram obtained from this hierarchical clustering includes all the colony candidates. A genetic algorithm is in charge of selecting the subset of colony candidates that optimizes the two performance metrics under study: the network communication time between users and applications, and the execution time of the algorithms that manage internally the placement of the applications in each colony. We implemented the NSGA-II, a common multi-objective approach for GAs, to evaluate our proposal. The results show that a meta-heuristic such as a GA improves the performance metrics by defining the fog colony layout through the use of the dendrogram. Nine different experiment scenarios, varying the number of applications and fog devices, were studied. In the worst of the cases, 137 generations were enough to the results of the GA dominated the solutions obtained with two control algorithms. The number of genetic solutions and their homogeneous distribution in the Pareto front were also satisfactory.



\end{abstract}


\end{frontmatter}



\section{Introduction}

Fog computing is a computing paradigm that extends the cloud to the devices located in the network infrastructure~\cite{bonomi2012fog}. These intermediate devices are usually called fog nodes or fog devices, and they own computational resources that allow them to store data or to execute applications. Because fog devices are geographically distributed and closer to users, fog computing reduces both the latency between users and applications, and the workload of the network infrastructure. This improvement benefits those applications with critical response times.

Although fog computing could be considered as a particular case of cloud computing that creates a continuum between the cloud and the users, there are marked differences between both of them \citep{DBLP:conf/fwc/MayerGSR17,moysiadis2018,GUERRERO2022101094}. The most important differences of fog computing, with regard to cloud computing, concern its much higher scale level, its wider geographical dispersion of the device locations, the more limited amount of resources in the devices, and the heterogeneity in the interconnection of the nodes. These features result in an important increase of the complexity in the management of the infrastructure. Previous research works have proposed the definition of fog colonies to reduce this complex management~\cite{SkarlatNSBL17}. Fog colonies are a subset of fog devices that acts such as a micro-data center, that aim to move from a centralized management of the infrastructure to a decentralized and isolated management for each colony.

The organization of the fog devices into fog colonies, that we call \textit{fog colony layout}, directly influences on the system performance~\cite{guerrero2018influence}. But, to the best of our knowledge, the number of papers that addressed the optimization of this fog colony layout is very limited.  

We propose to base the definition of the fog colony layout in the hierarchical clustering of the graph defined by the fog devices and their network connections. The result of the hierarchical clustering is a dendrogram, a tree structure in which each node represents a cluster of elements, i.e. a group of fog devices. We consider the set of all the nodes/clusters of the dendrogram as all the possible colony candidates. A subset of those clusters, constrained by an exhaustive and disjoint distribution of the devices into colonies, is selected to define the fog colony layout. This decision-making of selecting the colony candidates can undergo an optimization process. We consider the use of a genetic algorithm (GA), in particular a multi-objective version such as the NSGA-II, for this optimization process. 

Because one of the most important benefits of fog computing is to reduce the latency between users and applications, we are first interested in optimizing the network communication times between the users and the devices where the applications are allocated. And, because one of the most important drawbacks of the fog computing is the complexity in the management, we are secondly interested in optimizing the execution time of the algorithm that is in charge of managing the placement of the applications in the devices inside a colony. 

The implementation of the NSGA-II requires to adapt the operators and the solution representation to our specific optimization problem. On one hand, we represent the solutions as the selected clusters between the candidate suggested by the dendrogram. On the other hand, we adapt the traditional one-point crossover operator and the bit mutation operator to the case of a tree structure, such as the dendrogram. 

NSGA-II considers the dominance concept for the optimization process and, consequently, the result is a Pareto set of solutions instead of only one single solution. We compared the Pareto set obtained with the GA with the solutions obtained with two control algorithms. These control algorithms are the one that considers only one colony ---i.e., all the fog devices in the fog infrastructure are managed together--- and the one that creates the colonies with a pre-fixed size (number of devices). 

We evaluated nine different experiment scenarios, which where varied in terms of the number of applications and the number of fog nodes. The analysis of these experiments compared the solution of the GA, the Pareto front, with the solutions obtained with the two control algorithms, and it also analyzed the evolution of the solutions along the GA execution.

To sum up, the main contributions of this paper are:
\begin{itemize}
    \item The formal definition of the fog colony layout problem and its influence on the system performance.
    \item The use of the hierarchical clustering as a mechanism for the definition of the fog colony layout.
    \item The first genetic-based solution for the fog colony layout problem that adress the optimization of the network communication time and the execution time of the application placement management algorithm.
\end{itemize}

The paper is organized as follows: Section~\ref{sect_relatedwork} reviews related research works that also considered the use of fog colonies;  Section~\ref{sec_architecturedefinition} gives the details of the architecture we have defined to consider a dynamic organization of the devices into fog colonies; Section~\ref{sec_problemformulation} formally formulates the problem domain and the optimization we deal with; Section~\ref{sec_geneticoptimization} details the design decision of the implementation of the GA; Section~\ref{sect_experimentalevalutaion} presents the experiments and analyze the results; and finally, Section~\ref{sect_conclusions} summarizes the conclusions and establishes future research lines.

\section{Related work}
\label{sect_relatedwork}

The term of fog colony was initially coined by Skarlat et al.~\cite{SkarlatNSBL17}. This work defined a framework that supports a hierarchy of fog colonies with a head element in the cloud, i.e., the colony coordinator. Since this first definition of the fog colonies, many researches have considered this conceptual framework in their studies. There are even research efforts in the adaptation of technologies to this fog landscape~\cite{9647666,SkarlatKRB018,OGUNDOYIN2022108942} and in the implementation of the framework in real scenarios~\cite{10.1007/978-3-030-35653-8_23}.


Optimization processes can be applied to almost all the components of a system. In the case of fog colonies, a literature review shows that most of the optimization proposals are focused on the problem of service placement, mapping application/service instances to fog colonies and devices. A wide range of optimization algorithms have been studied for the optimization of this fog service placement problem, such as: greedy algorithms~\cite{10.7717/peerj-cs.588, azizi2019qos, 8589492,azizi2019b,8710172}, search algorithms~\cite{8710172,9706284, Nicopolitidis2019}, integer linear programming~\cite{9706284}, analytical resolution~\cite{8108080,Tran2019a}, benchmarking~\cite{venticinque2019methodology}, multi-dimensional optimization~\cite{9119416}, self-adapting function~\cite{Ghaferi2022a}, genetic algorithms~\cite{10.7717/peerj-cs.588}, evolutionary optimization~\cite{9685175,10.1002/spe.2939}, cuckoo search algorithm~\cite{Liu2022a}, particle swarm optimization~\cite{doi:10.1080/08839514.2021.2008149}, imperialist competitive algorithm~\cite{10.1002/spe.3111}, gray wolf optimization~\cite{10.1002/spe.2986}, or fuzzy logic~\cite{Tavousi2022a}. Other optimization scopes have been also considered, but in a more limited number of studies: load balancing~\cite{10.1145/3506718,Refaat2019}, scheduling~\cite{9315281,HOSSAIN2021102336}, offloading~\cite{9382907,VAKILIAN2022103428}, resource provisioning~\cite{ETEMADI2020109,10.1007/978-981-16-6723-7_33}, monitoring resources~\cite{8944166}, and autoscaling~\cite{10.1007/978-3-030-86162-9_28}.

By the analysis of the experimental phases of those studies, we conclude that the definition of the fog colony layout is usually performed manually by the infrastructure administrator~\cite{tran2020}, using criteria related with the geographical distribution of the fog devices or the size of the fog colonies. Although there are enough evidences that the organization of the layout of the fog colonies directly influences on the performance of the system~\cite{guerrero2018influence,NIKOLOPOULOS2022100549}, the number of studies that pay attention to the optimization or the autonomous organization of the fog colony layout is very limited. 

In terms of the autonomous organization, Nikolopoulos et al.~\cite{iot3010005} proposed a framework for a self-configured, self-adapted and context awareness organization of the fog nodes into fog colonies. They used a publish/subscribe protocol to allow fog nodes to become actively aware of their operating context, share contextual information and exchange operational policies. They explored different methods of communication between the fog nodes, but they did not deal with the optimization of the fog colony layout, and its influence in the system performance. Similarly, Lordan et al.~\cite{10.1007/978-3-030-85665-6_17} defined a framework which allows to fog nodes to establish relations with other nodes to create new colonies or to participate into already-existing ones through the use of an API. But this work neither deals with establishing a fog colony organization to optimize the system.

In terms of the optimization of the fog layout, we previously explored this problem by proposing to use centrality indexes to define the fog coordinator candidates and, once the coordinator candidates are selected, each fog devices is attached to the closest fog coordinator, creating the fog colonies~\cite{guerrero2018influence}. Although this process is semi-automatic, the action of the infrastructure administrator is still required to determine the size of the fog colonies in the process of selecting the fog coordinators.  

Tran et al.~\cite{tran2020} stated that, between other design features, fog infrastructures should be able to deal with the scalability of the system. They propose a model to allow automatic addition of new fog nodes into fog colonies. Concretely, the newly added fog node identifies the most suitable colony by detecting its geographical position.

Finally, Hatti et al.~\cite{HATTI202161} proposed to automatically determine the fog colonies by clustering the fog devices. Colonies are formed by clustering the devices located in the geographical area using a k-mean based algorithm~\cite{8241329}. The algorithm uses the Euclidean distance to group the devices into colonies and the nearest devices are initialized as the centroids of the clusters.


Our proposal differs from the related research in that is the first work that combines genetic optimization and graph partitioning techniques for a non human-assisted definition of the fog colony layout. This colony definition is addressed to optimize the performance of the system by searching a trade-off to both reduce the service placement time and the network times. 

Note that the use of complex network features is a common solution to deal with the optimization of general distributed systems but also for the particular case of fog computing~\cite{lera2018availability,Azimzadeh2022,guerrero2020optimization}. But, to the best of our knowledge, our work is the first one that applies a hierarchical clustering for the organization of the fog colony layout.


\section{Architecture definition}
\label{sec_architecturedefinition}

The definition of the architecture is inspired in the work of Skarlat et al.~\cite{SkarlatNSBL17} and it enables a conceptual organization of the computing resources into a layered distribution of responsibilities both with regard to the execution of applications and the management of the infrastructure.

Fog computing is a computing paradigm that can be understood as an extension of cloud architecture creating a continuum between the cloud and the users~\cite{bonomi2012fog}. In-network devices,  located between the data center and the clients, include computational and storage resources, allowing them to store/process data, execute applications, etc. These intermediate elements of the infrastructure are known as fog nodes or fog devices. Fog devices are closer to the users and geographically distributed, reducing the latency between users and application/data location. 

Our framework organizes the fog devices into groups, known as fog colonies. These colonies are formed by one fog coordinator and a set of additional fog devices. The fog coordinator is just a fog device that has activated the coordination module when required. Each fog colony has one and only one fog coordinator, and one fog device belongs to one and only one fog colony. Fog devices, and accordingly fog coordinators, are able to deploy and allocate applications.

\begin{figure}
	\centering
	\includegraphics[width=\textwidth]{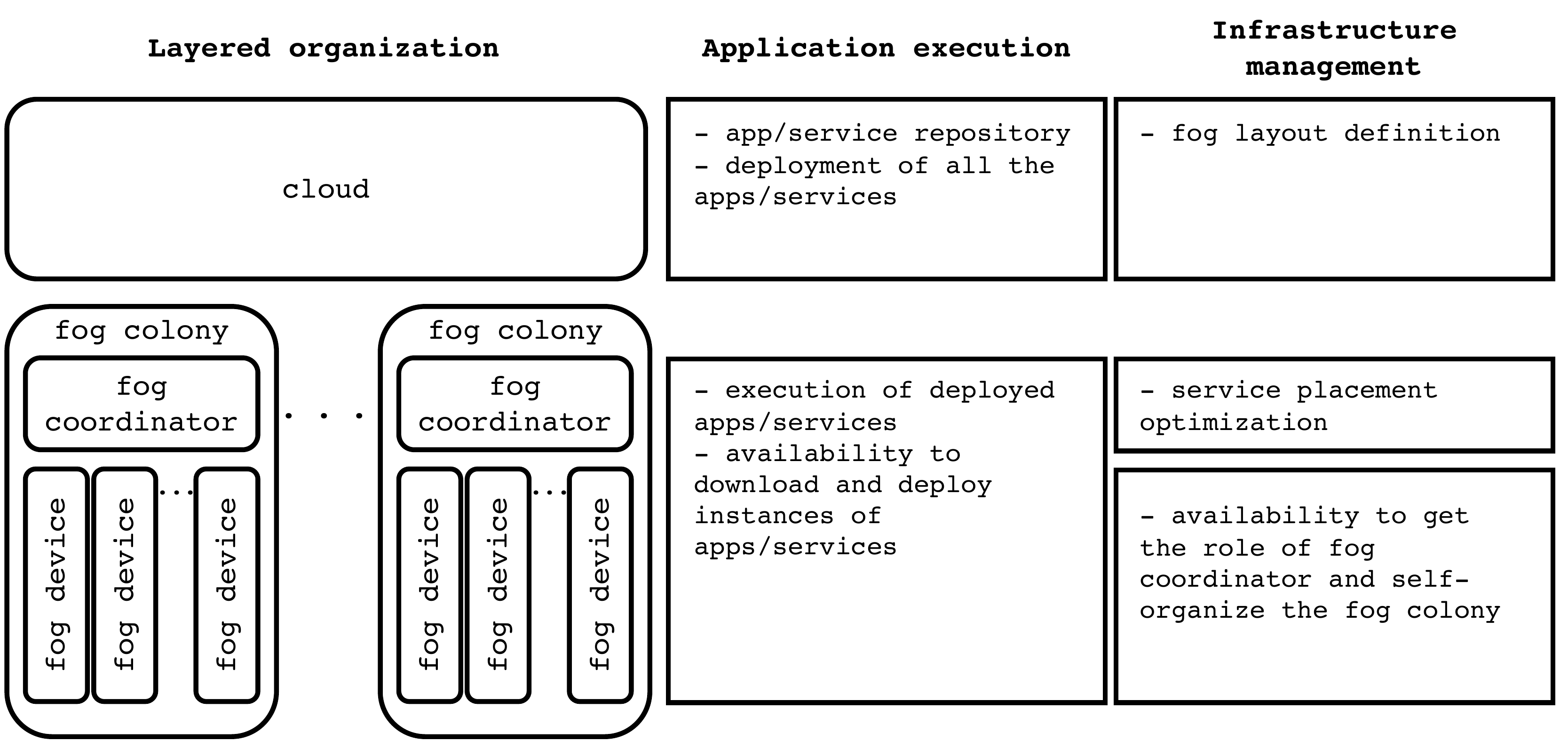}
	\caption{Layered definition of the infrastructure.} \label{layered_infrastructure}
\end{figure}

In summary, the proposed layered infrastructure is divided into three levels, cloud, fog coordinator and fog devices, corresponding these two last to the fog colonies. The roles of the application execution and infrastructure management are distributed between these three layers (\figurename{~\ref{layered_infrastructure}}).

We consider that applications follow a multi-service pattern that is increasingly being used in fog applications. Applications are deployed as a set of services that mutually interoperate to accomplish the task implemented by the application~\cite{brogi2020place,WANG2022103354}. Accordingly, services can easily be scaled up, by downloading its encapsulating element and executing it, or scaled down, by simply stopping and removing instances of the service.

In our conceptual framework, the cloud acts as a central repository for all the services and, additionally, it also executes instances of all the applications in the infrastructure. Thus, if an application was not deployed in any fog devices, the user requests would be served by the cloud. When a fog device is designated to deploy a service, it downloads the encapsulating element of the service (the image including the service source code and data) from the cloud, and starts its the execution. Fog coordinator behaves the same as fog devices for the case of application execution.

In the case of infrastructure management, the cloud is in charge of defining the fog colony layout. Once that the fog layout is determined and the fog coordinators are selected, these are notified. Any fog device is able to act as a fog coordinator when required. 

Consequently, the fog colony layout can be self-managed and dynamically adapted to the changing conditions of the system. The cloud just needs to periodically evaluate the colony layout and to decide if changes in the fog colonies are required. 

The tasks of the fog coordinator are to manage the colony, for example by optimizing the placement of the applications inside the colony, and to be in charge of the routing of the communications between colonies. The fog coordinator is aware of the fog devices included in its own colony, and which the coordinators of the other colonies are. The details of the communication between fog devices are explained in the following section (Section~\ref{sec_communicationpatterns}).

The placement of the services of one application is not constrained by the colony layout. The set of services of an application can be either deployed in the same fog device, in devices of the same fog colony, or, even, devices from different colonies. The optimization algorithm executed by the fog coordinator will map service instances and devices inside its own colony. The services that were not deployed in the colony would be requested to other colonies or to the cloud.

\subsection{Communication patterns}
\label{sec_communicationpatterns}

Following the architecture defined by Skarlat et al.~\cite{SkarlatNSBL17}, we establish the communication routing between devices under the assumption that a fog device can only communicate with the fog devices in its own colony. In that case, the communication time for services inside a colony is the sum of the latencies of the shortest path between the devices that place the services.  

On the contrary, if it is necessary to communicate with devices in other colonies, or with the cloud, the communication is performed via the colony coordinators~\cite{SkarlatNSBL17}. In those cases, the communication time is calculated as the summation of (i) the shortest path between the fog device with the origin service and the coordinator of its own colony, (ii) the shortest path between the coordinator devices of both colonies, and (iii) the shortest path between the coordinator of the neighbour colony and the device that places the target service.

\begin{figure}[t!]
\centering
\subfloat[Communication time for an application with services placed in a neighbour colonies.\label{communicationtimes_2col}]{\includegraphics[width=0.75\textwidth]{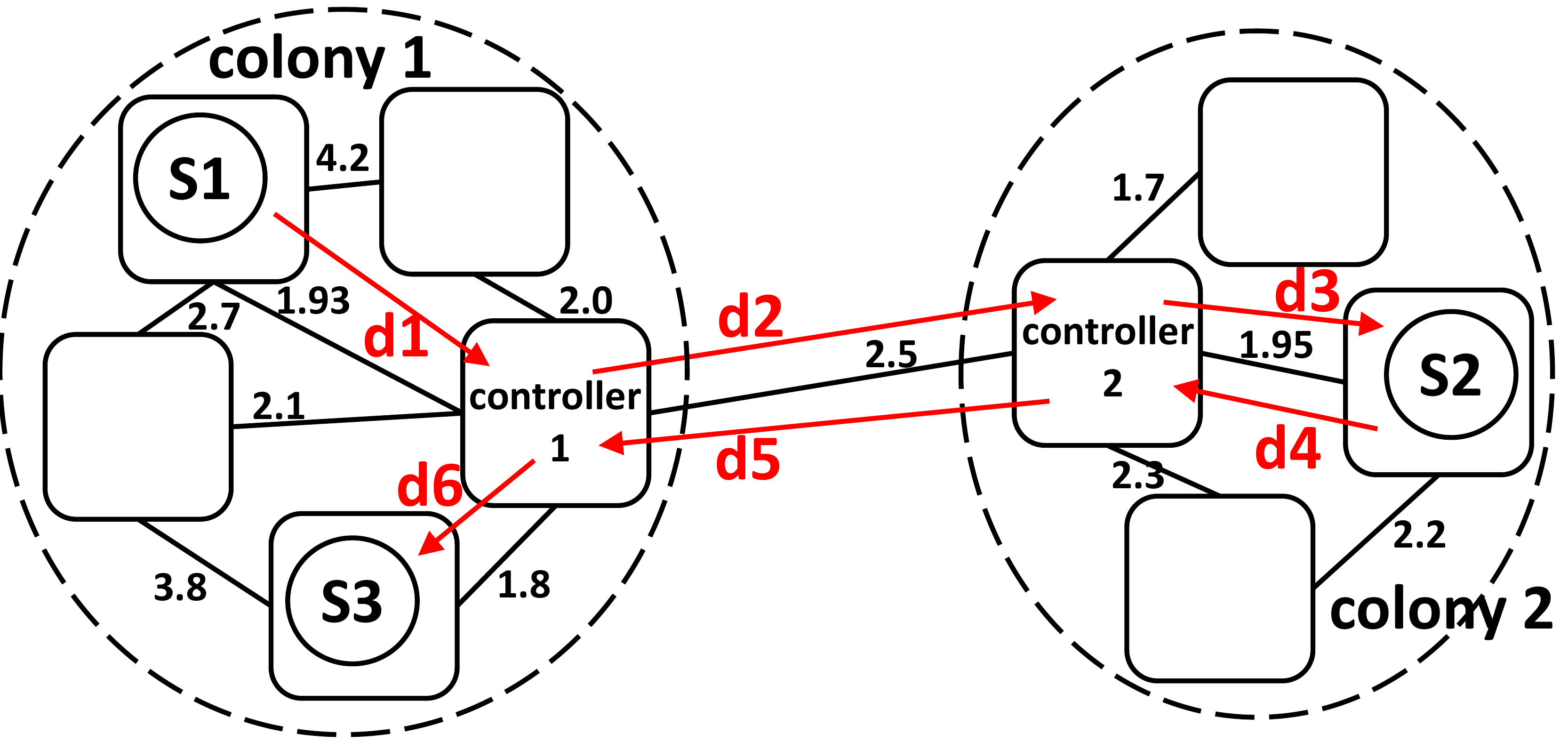}}
\\
\subfloat[Communication time for an application with services placed in only one colony.\label{communicationtimes_1col}]{\includegraphics[width=0.75\textwidth]{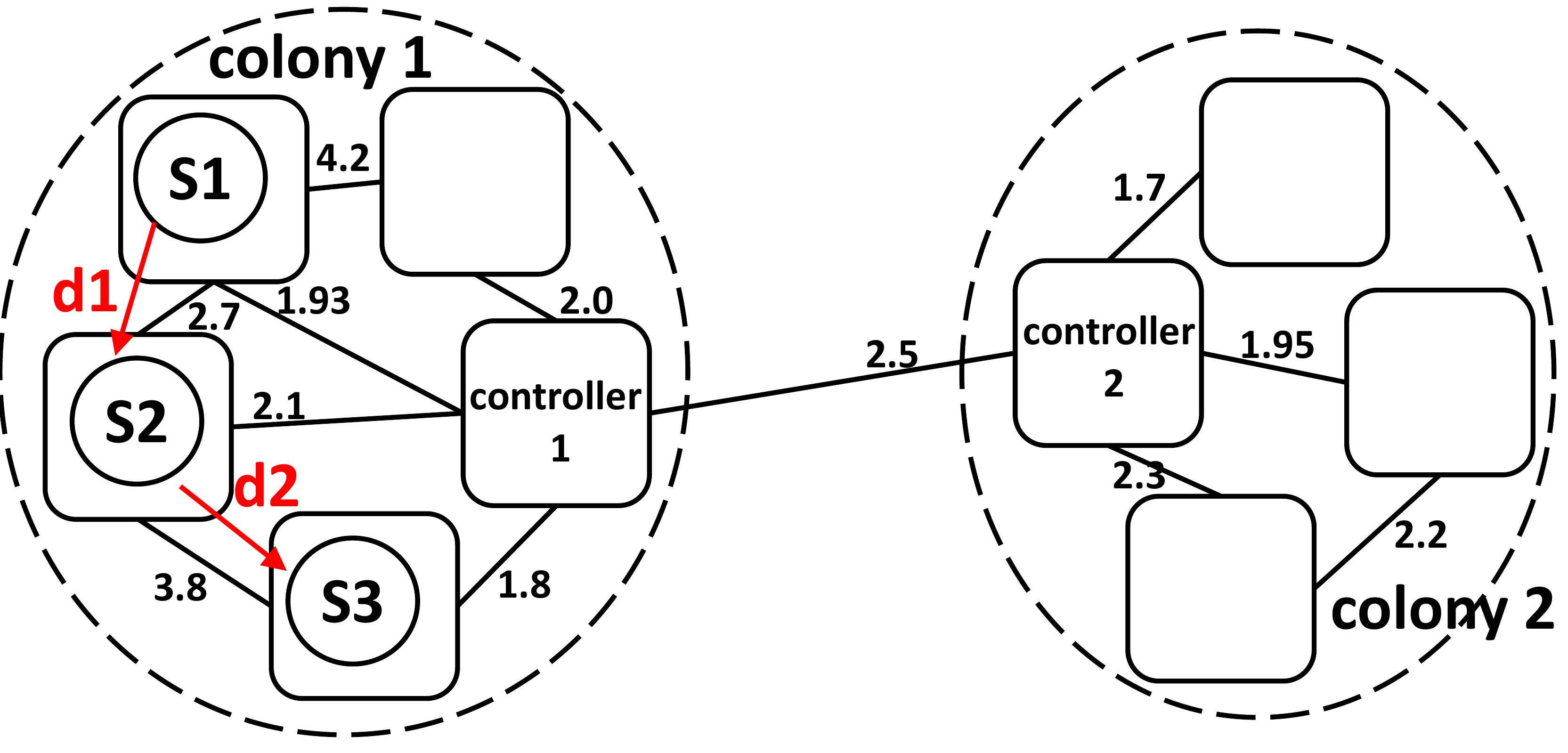}}
\caption{Examples of communication times with regard to the sevice location.}
\label{fig_communicationtimes}
\end{figure}

An example of this second case is shown in \figurename{~\ref{communicationtimes_2col}}. Consider an application of three services which are executed sequentially. The first and last services (S1 and S2) are deployed in the same colony. The second service is allocated in the neighbour colony. By this, two inter colony communications (d2 and d5) and 4 intra colony communications are required. The time sequence of the communication delays is $d1 + d2 + d3 + d4 + d5 + d6$, resulting in a total network delay of $1.93+2.5+1.95+1.95+2.5+1.8=12.63$ time units. On the contrary, if the three services of the application had been placed in the same colony, the communication path would have been reduced to only two inter colony communications. This is illustrated in  \figurename{~\ref{communicationtimes_1col}}, resulting the delays $d1 + d2$ in a total time of $2.7+3.8=6.5$ time units.



The same assumption for routing requests between services is also established for the requests of the users. When the user and the service are in the same colony, the user requests are routed through the shortest path between the device where the user is connected to and the devices that places the service. If the services is placed in a different colony, the user request is routed via the fog coordinators.

\section{Problem formulation}
\label{sec_problemformulation}

The model considered in this paper consists of fog devices, fog colonies, interconnection networks, applications composed of services, and users. First, we consider a fog device as any device in the infrastructure that has computational resources that allow it to execute instances of a service. Denote fog nodes as $f_n \in F$. Fog devices are characterized by the resource capacity, $f_n[rc]$. Without loss of generality, we use a scalar value $rc$ to represent the resource capacity of the devices, but it could be extended straightforwardly to a n-tuple which models the capacity of each single resource in the device, for example, processor, memory, disk, etc. 

The fog nodes are connected through a network. A network link $n_{f_n,f_m} \in N$ represents a network connection between nodes $f_n$ and $f_m$ that does not go through any other fog device, i.e., a direct network connection between the fog devices. Network links are characterized by the latency of transmitting data across this connection, $n_{f_n,f_m}[lat]$.

The fog infrastructure $I$ is the set of fog devices $F$ interconnected by the set of networks links $N$. It can be modelled as the graph $I=(F,N)$ where the the fog devices are nodes and the network links are edges. 

As it is explained in Section~\ref{sec_architecturedefinition}, fog devices are organized into fog colonies. A fog colony is a disjoint subgraph of the fog infrastructure $I$. Since the fog infrastructure is defined as a graph, we are able to obtain its community structure. We define the fog colonies of the infrastructure through the dendrogram of the fog devices, that is the output obtained from a hierarchical clustering method.

A dendrogram is a hierarchical binary cluster tree that is commonly used to organize the elements into clusters~\cite{murtagh2012algorithms}. The hierarchical clustering can be designed as agglomerative (bottom-up)\footnote{The algorithm starts considering each element as a cluster of one element. Iterative, two clusters are merged as the algorithm moves up the hierarchy, until only on cluster is obtained.} or divisive (top-down)\footnote{The algorithm starts considering one cluster with all the elements. Iterative, one cluster is selected and split as the algorithm moves down the hierarchy. The algorithms finishes when all the clusters are one element in size.} process. In both cases, the sequential evolution of the algorithm is reflected in the dendrogram on the height where the split/join of two clusters is represented.


We propose to use the dendrogram to represent all the possible fog colonies' structures in the infrastructure. Each node of the dendrogram corresponds to a candidate cluster to organize the fog devices into fog colonies. For example, the leaf nodes corresponds to the fog colonies (clusters) that only include one fog device (element)\footnote{Fog colony/cluster and fog device/element are used indistinctly in the paper. The first name refers to the field of the fog infrastructure and the second one to the field of the hierarchical clustering.}, the root node to the fog colony that include all the fog devices, and the other nodes are intermediate cases. We identify each candidate fog colony as $C_i \in \mathbb{C}$ and, consequently, the corresponding node of the dendrogram can be also identified with the same label $C_i$. Each of these $C_i$ corresponds to a subset of fog devices, the ones that this cluster includes, $C_i \subset F$. Additionally, each fog colony is also characterized by the colony coordinator, $f_{coord}^{C_i} \in C_i$, the fog device in the colony in charge of managing the own colony (Section~\ref{sec_architecturedefinition}). The colony coordinator is defined by choosing the device with the highest centrality index in the colony, to reduce the communication distances~\cite{lera2018availability}.

The dendrogram represents all the candidate fog colonies, through its nodes, but an specific fog colony layout, $L_x \in \mathbb{L}$, is defined as a subset of those nodes (or candidate fog colonies), $L_x \subset \mathbb{C}$. Not all the possible subsets of clusters correspond to suitable colony layouts, being constrained by an exhaustive and disjoint distribution of the devices between the colonies. A proper colony layout includes all the fog devices in one and only one cluster ($\exists ! C_i \in L_x: f_n \in C_i$), i.e., fog devices cannot be included in more than one cluster, $\bigcap_{C_i \in L_x} C_i= \O$, and all the devices must be included in a fog colony, $\bigcup_{C_i \in L_x} C_i= F$. The set of proper colony layouts forms the colony solution space $\mathbb{L}$.

\begin{figure}[t!]
	\centering
	\includegraphics[width=\textwidth]{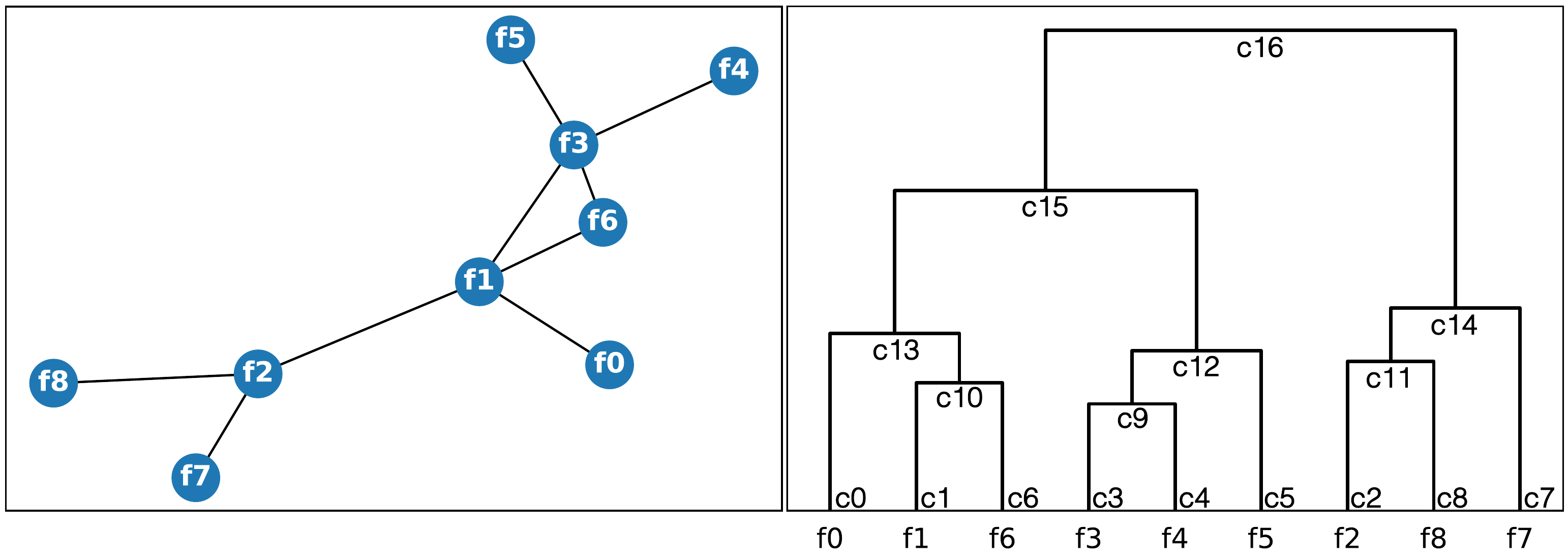}
	\caption{Example of fog infrastructure and its dendrogram.}\label{fogdendrogram}
\end{figure}

Figure~\ref{fogdendrogram} shows the example of a fog infrastructure modelled with a graph and the dendrogram obtained from applying a hierarchical clustering. In this example, the infrastructure is composed by 9 fog devices ($F=\{f_0..f_{8}\}$) and there are 17 candidates of fog colonies ($\mathbb{C}=\{C_0..C_{16}\}$). The resulting fog colony layout has to include all the fog devices without intersections. For example, a  suitable layout would be obtained by selecting the clusters $L_{ex}=\{C_{13}, C_9, C_5, C_{14}\}$ which results in the fog colonies formed, respectively, by the devices $C_{13}=\{f_0, f_1, f_6\}$, $C_9=\{f_3, f_4\}$, $C_5=\{f_5\}$, and $C_{14}=\{f_2, f_8, f_7\}$. Note that all the devices are included in one and only one colony.  On the contrary, if we select clusters $L_{ex'}=\{C_{15}, C_{10}, C_{11}\}$, the resulting fog colonies are $C_{15}=\{f_0, f_1, f_6, f_3, f_4, f_5\}$, $C_{10}=\{f_1, f_6\}$, and $C_{11}=\{f_2, f_8\}$, which does not satisfied any of both constraints to obtain a proper layout: the layout is not disjoint because devices $f_1$ and $f_6$ are included in two fog colonies; the layout is not exhaustive because device $f_7$ is not included in any colony. This second example should be transformed in a suitable layout with a agglomerative or divisive process. The agglomerative solution reduces the number of fog colonies, resulting in $L_{agglomerative}=\{C_{15}, C_{14}\}$ in our example. The divisive solution increases the number of colonies, with $L_{divisive}=\{C_{0}, C_{10}, C_{12}, C_{11}, C_{7}\}$ for the example.

Note that not all the combinations of fog devices subsets are considered as candidate colonies. This is based on the idea that the colonies should be formed by neighboring devices, in fact, this is the criteria of the hierarchical clustering to define the clusters. For example, we can see in Figure~\ref{fogdendrogram} that the colony $\{f_7, f_4\}$ is not considered as candidate and these two devices are only in the same colony for the case of the \textit{root} candidate ($C_{16}$), which includes all the devices. This is reasonable because the network distance between these two devices is equal to the diameter of the graph (the greatest distance between two vertices in the graph).

The fog applications, $A_o \in \mathbb{A}$ are composed of a set services, $s_p \in S$ interrelated through a request/data relationship, $r_q \in R, r=(s_p,s_{p'})$, and modelled as a graph $A_o=(S,R)$. Services are characterized by the resource requirements, $s_p[rr]$, modeled with a scalar value $rr$, as established in the definition of the fog devices.

These applications are requested by users and \textit{things}\footnote{Things are entities or physical objects (end devices) that have the ability to request the execution of application to obtain a service or to process the data they generate}, $u_j \in U$. These users/things are defined as $u_j[A_o, f_n, \lambda]$, where $A_o$ is the application that the user requests, $f_n$ is the fog device in which the user is connected to, and $\lambda$ is the rate in which the application is requested. For the shake of simplicity, we consider that users only request one application. This can be easily extended by mapping real users with as many users as applications they request.

Each service $s_p \in S$ has to be placed on a fog device $f_n \in F$ to execute the application and to respond the user requests. The same service can be placed in several fog devices, and these copies of the service are named instances, $s_p^q$. The placement of the services into the devices is represented with a binary matrix $P$ where the rows represent the services and the columns the fog devices. An element $p_{s_p,f_n}=1$ if an instance of service $s_p$ is placed in device $f_n$, and $p_{s_p,f_n}=0$ if the service is not deployed in the device. Remember that the layout of the infrastructure is based in organizing the fog devices into fog colonies and that the placement problem is not addressed globally, and it is handle locally in each for colony. Thus, instead a global placement matrix, we have defined placement matrices for each colony, $P^{C_i}\ \forall C_i \in L_x$. Theses placement matrices are smaller than the global placement matrix and, consequently, the scale of the placement optimization is reduced to lower computational costly processes.

The placement is constrained by the resource capacity of the fog devices. The sum of the resource requirements of all the services placed in a device has to be smaller than the resource capacity of the device, $(\sum_{s_p \in S} p_{s_p,f_n} \times s_p[rr] )\leq f_n[rc]$.

In summary, this paper addresses the problem of finding:
\begin{gather}
L_x, \label{eq_def_layout}\\
P^{C_i}\ \forall C_i \in L_x \label{eq_def_place}
\end{gather}
where the definition of Eq.~(\ref{eq_def_layout}) is constrained by:
\begin{gather}
\bigcup_{C_i \in L_x} C_i= F,\\
\bigcap_{C_i \in L_x} C_i= \emptyset
\end{gather}
and the definition of Eq.~(\ref{eq_def_place}) is constrained by:
\begin{gather}
\left(\sum_{s_p \in S} p_{s_p,f_n} \times s_p[rr] \right)\leq f_n[rc], \forall f_n \in F. 
\end{gather}

\subsection{Optimization definition}

Our objective is to find the definitions of Eq.~(\ref{eq_def_layout}) and Eq.~(\ref{eq_def_place}) that optimize the overall execution time for the algorithm that places the services in the devices in a colony ($placement\_time$), and that optimize the overall response time of the applications ($response\_time$). 

The optimization of the fog service placement is a NP-hard problem directly influenced by the number of application instances and the number of fog devices involved in the placement process~\cite{GUERRERO2022101094}. If the devices in the infrastructure are grouped into fog colonies, and an independent placement process is executed for each colony, the size of the problem is reduced with regard to considering the placement into all the infrastructure. Consequently, it is clear that the definition of Eq.~(\ref{eq_def_layout}) will influence in the $placement\_time$ and it will be minimized as smaller colonies are considered.

The placement algorithm is executed independently in each colony by the colony coordinator and, consequently, these executions can be parallelized.  We propose to use the average of the execution time of the placement algorithm in each of the colonies as the indicator of the $placement\_time$. We formally defined the $placement\_time$ as:
\begin{gather}
placement\_time = \frac{ \sum_{C_i \in L_x} placement\_time_{C_i}}{|L_x|}
\end{gather}
where $placement\_time_{C_i}$ is the specific execution time of the placement for the colony $C_i$. Due to the complexity to determine this time in an analytical way, we propose to measure it as the real execution time of the placement. This is a common solution when emulation or simulation is required to evaluate the fitness function of an optimization process~\cite{10.1002/spe.2631}. In our particular case, this does not suppose an increment in the optimization complexity because the placement of the services has to be performed to evaluate the second optimization objective anyhow, the overall response time of the applications ($response\_time$).

The response time of an application for a given user, $response\_time_{A_o,u_j}$ is determined by the execution times of each service and by the latency times of the communication links between the devices that place the services: 
\begin{gather}
response\_time_{A_o,u_j} = execution\_time_{A_o,u_j} + network\_time_{A_o,u_j}
\end{gather}
Without loss of generality, and to reduce the influence of the fog device resource heterogeneity, we consider that $execution\_time_{A_o,u_j}=0$, i.e., we are really only evaluating the network time component of the response time. Our optimization process is focused on the fog colony layout, and this layout influences in the network time but not in the execution time of the services. By this, we prefer to eliminate the influence that the device heterogeneity would have in the experimentation phase of our proposal. Anyhow, the model could be easily adapted by including the execution speed of the devices and the required execution time by the services.

Remenber that our architecture considers that applications are composed of a set of services modelled as a graph $A_o=(S,R)$. Accordingly, the $network\_time_{A_o,u_j}$ can be calculated as the sum of the single network times for each pair of related services $r_q \in R, r_q=(s_p,s_{p'})$ of the executed application:

\begin{gather}
network\_time_{A_o,u_j} = network\_time_{u_j,s_{root}} + \sum_{r_q \in R_{A_o}}network\_time_{s_p,s_{p'}}
\end{gather}

Let define the pair of \textit{source-target devices} as the pair of fog devices that requires to establish a network connection because either the source device has connected a user that requests a service that the target device allocates, or the source device allocates a service that requests a second service placed in the target device. Remember that the direct communication routing between \textit{source-target devices} is only able when the devices are in the same colony. For those cases with devices in different colonies, the communication is established through the fog coordinators (Section~\ref{sec_communicationpatterns}). Accordingly, when the \textit{source-target devices} are placed in the same colony, the $network\_time_{s_p,s_{p'}}$ is determined as the network latency of the shortest path between them ($shortestLat$). But, when these devices are located in different colonies, the $network\_time_{s_p,s_{p'}}$ is determined by the closest neighbour ($closestColony$), that is the sum of latencies of the shortest network paths ($shortestLat$) for the pairs: source device-source colony coordinator; source colony coordinator-neighbouring colony coordinator; and, neighbouring colony coordinator-target device. The neighbouring colony is determined as the one with the shortest latency between its own coordinator and the source colony coordinator, between all the neighboring colonies that have the requested service placed in some of its devices. The cloud provider acts as the neighbouring colony either when its network distance is smaller than the distance with the colony, or when no other colony allocates the requested service. In that specific case, the $network\_time_{s_p,s_{p'}}$ is sum of the latencies of the shortest paths of both source device-source colony coordinator and source colony coordinator-cloud ($cloudLat$).
Thus the response time for a single user is formally defined as:


{\footnotesize
\begin{gather}   network\_time_{s_p,s_{p'}}=
\left\{
\begin{array}{ll}
shortestLat(f_n,f_{n'}) & f_n \in C_i,\ \exists f_{n'} \in C_i \\
      & \ |\ p_{s_p,f_n}=p_{s_{p'},f_{n'}}=1 \\
&\\
      \min \left( closestColony(s_{p},s_{p'}), cloudLat(s_p,s_{p'}) \right) & otherwise \\
\end{array} 
\right.
\end{gather}
}
where
{\footnotesize
\begin{gather}
\begin{array}{l}
closestColony(s_p,s_{p'}) =  shortestLat(f_n,f^{C_i}_{coord}) + \\ 
\\
 + \min \left( shortestLat(f^{C_i}_{coord},f^{C_{i'}}_{coord}) + shortestLat(f^{C_{i'}}_{coord},f_{n'}), \forall C_{i'}\ |\ \exists f_{n'} \in C_{i'} \wedge p_{s_{p'},f_{n'}}=1 \right)
\end{array}
\end{gather}
}
and
{\footnotesize
\begin{gather}
\begin{array}{ll}
cloudLat(s_p,s_{p'}) = & shortestLat(f_n,f^{C_i}_{coord}) + \\ 
&\\
& + shortestLat(f^{C_i}_{coord},Cloud)
\end{array}
\end{gather}
}

The $network\_time_{u_j,s_{root}}$ is analogous defined to $network\_time_{s_p,s_{p'}}$, considering the network time between the user and the root service of the application.

Finally, the overall response time for all the users in the infrastructure, $response\_time$, is defined as the mean value of the single response times for all the users:
\begin{gather}
response\_time = \frac{\sum_{u_j \in U} response\_time_{A_o,u_j} }{|U|}
\label{eq_reponsetimeobjective}
\end{gather}

Note that smaller colonies have less resources to allocate services because they are composed by a smaller number of devices. Consequently, with smaller colonies, shifting the placement of the services to other colonies or to the cloud is more likely. On those cases, the service requested by an user will not be placed in the colony where the user is connected to and the $response\_time$ will be increased due to the networking time between colony coordinators~\cite{SkarlatNSBL17}. On the contrary, bigger colonies are able to allocate more services but for those cases that the services are shifted to other colonies, the network coordination time between coordinators is larger because the coordinators are further away~\cite{guerrero2018influence}.

In summary, the colony layout influences on the $response\_time$ and the $placement\_time$, with opposed effects when the colony sizes are increased or decreased. This is a common multi-objective optimization problem that cannot be solved analytically because of the size of the problem, and that requires some meta-heuristic to be solved~\cite{GUERRERO2022101094}. Consequently, the optimization problem is formulated as:
\begin{eqnarray}
\text{Define }& L_x, \label{eq_def_layout2}\\
&P^{C_i}\ \forall C_i \in L_x \label{eq_def_place2}\\
\text{to minimize } &response\_time \\ 
& placement\_time
\end{eqnarray}

We propose to use a GA for solving this multi-objective optimization problem. In particular, we have implemented a modified version of a NSGA-II (Elitist Non-Dominated Sorting Genetic Algorithm)~\cite{deb2002fast}, which evaluates the fitness functions through the emulation of the placement process of the services into the colonies. The next section describe the details of the implemented NSGA-II algorithm. 




\section{Genetic optimization of fog colony layout}
\label{sec_geneticoptimization}

We propose a modified version of NSGA-II~\cite{deb2002fast} for our optimization problem. NSGA-II is a Pareto-based multi-objective genetic algorithm that has shown high performance for problems with two and three optimization objectives.

GAs are meta-heuristic search algorithms that work with a population, which corresponds to a set of solutions to the problem to be optimized. GAs improve this population through an evolution process based on the stochastic combination of the best solutions \cite{holland1992adaptation}. The solutions of a GA evolve along generations limited by a finish condition. Each new generation is created by selecting, combining, and modifying solutions from the previous generation, creating an offspring of the same size (number of solutions) than the population size.

NSGA-II mainly differs from traditional single objective GAs in how the goodness of the solutions is calculated to be compared between them. NSGA-II is a Pareto-based algorithm that uses the concept of dominance to order the solutions~\footnote{Solution $s_1$ non-dominates solution $s_2$, if $s_2$ is better for at least one objective.}. NSGA-II recursively classifies the solutions in fronts, which are created with the solutions that are not included in previous fronts and that are not dominated by any other solution. Solutions in the former fronts have higher goodness than solutions in latter fronts. Solutions inside the fronts are ordered by the crowding distance, that is calculated as the minimum Euclidean distance of the objectives values from one solution to the others. NSGA-II considers that dispersed solutions are more significant than the solutions that are concentrated together.

Algorithm~\ref{nsga2} shows the baseline structure of the NSGA-II. The quality of a GA implementation is directly influenced by the solution representation, crossover and mutation operators, offspring selection, fitness function, selection operator,  and parameters calibration~\cite{larranaga1999genetic}. NSGA-II defines some of those elements by default (elitism for offspring selection and binary tournament for selection operator), but others need special attention.

\begin{algorithm}[t!]
	\caption{Multi-objective genetic optimization algorithm~\cite{deb2002fast}}
	\label{nsga2}
	\begin{algorithmic}[1]
		\Procedure{NSGA-II}{}
		\State $P_t \gets generateRandomPopulation(pop_{size})$ \label{alg_rand_gen}
		\State $placement,fitness.placement\_time \gets servicePlacement(P_t)$
		\State $fitness.response\_time \gets calculateNetworkDistance(P_t,placement)$
		\State $fronts \gets calculateFronts(P_t,fitness)$
		\State $distances \gets calculateCrowding(P_t,fronts,fitness)$
		\For{$i\ $in$\ 1..gen_{num}$}
		\State $P_{off} = \emptyset$ 
		\For{$j\ $in$\ 1..pop_{size}$}
		\label{alg_sel1}
		\If {$random(0,1) < \rho_{cross}$} 
		\State $father1 \gets binaryTournament(P_t,fronts,distances)$
		\State $father2 \gets binaryTournament(P_t,fronts,distances)$ 
		\label{alg_sel2}
		\State $child1,child2 \gets crossover(father1,father2)$
		\EndIf
		\If {$random(0,1) < \rho_{mut}$} 
		\State $rndOp \gets random(0,1)$
		\If {$rndOp < \rho_{mut}^{join}$} 
		\State $joinMutation(child1)$,$joinMutation(child2)$
        \ElsIf {$\rho_{mut}^{join} > rndOp < \rho_{mut}^{split}$} 
        \State $splitMutation(child1)$,$splitMutation(child2)$
		\EndIf
		\EndIf
		\If {$random(0,1) < \rho_{rep}^{agg}$} 
		\State $repairAgglomerative(child1)$,$repairAgglomerative(child2)$
		\Else
		\State $repairDivisive(child1)$,$repairDivisive(child2)$		
		\EndIf
		\State $P_{off} = P_{off} \cup \{child1, child2\} $ 
		\EndFor \label{alg_end_for}
		\State $placement,fitness.placement\_time \gets servicePlacement(P_{off})$
		\State $fitness.response\_time \gets calculateNetworkDistance(P_{off},placement)$
		\State $P_{union} = P_{off} \cup P_{t}$  \label{alg_join}
		\State $fronts \gets calculateFronts(P_{union},fitness)$ 
		\State $distances \gets calculateCrowding(P_{union},fronts,fitness)$
		\State $P_{union} = orderElements(P_{union},fronts,distances)$ \label{alg_order}
		\State $P_{t} = P_{union}[1..pop_{size}]$  \label{alg_half}
		
		\EndFor
		\State $Solution = fronts[1]$ \#the Pareto front
		
		\EndProcedure
	\end{algorithmic}
\end{algorithm}

		
		
\begin{figure}[h!]
	\centering
	\includegraphics[width=\textwidth]{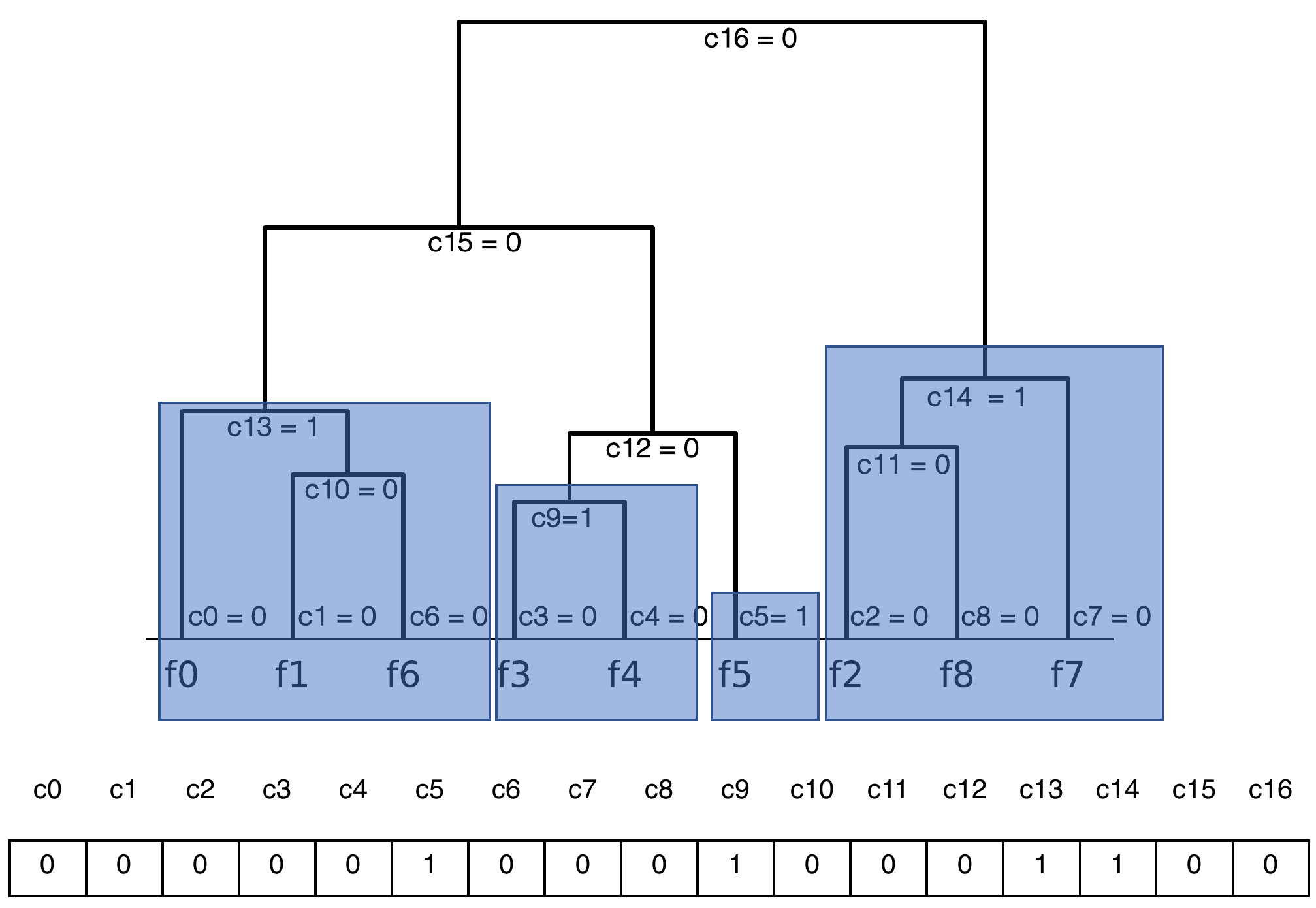}
	\caption{Example of solution representation and its chromosome.}\label{chromosome}
\end{figure}

Chromosomes are the solution representations in a GA. They are commonly modeled with an array or a matrix. In our particular case, the genetic optimization searches a suitable fog colony layout. As we explained in Section~\ref{sec_problemformulation}, the colony layout is defined through the dendrogram, particularly, the selected candidate fog colonies. Accordingly, we represent the solutions using a dendrogram with binary labeled nodes. For internal storage and processing, this dendrogram is transformed into a binary array chromosome of size the number of candidate fog colonies ($|\mathbb{C}|$), where each position of the array represents a candidate colony (a node of dendrogram), and the binary value of the position indicates if the candidate colony is selected (value 1) or not (value 0). For example, considering $L_{ex}=\{C_{13}, C_9, C_5, C_{14}\}$ from the case in Figure~\ref{fogdendrogram}, the chromosome that represents this solution is $[0,0,0,0,0,1,0,0,0,1,0,0,0,1,1,0,0]$ and the corresponding binary labeled dendrogram is represented in \figurename{~\ref{chromosome}}.  Remember that not all the combinations of selected fog colonies are valid solutions for the problem. It is also important to remark that, although the solutions are represented with a binary array, the crossover and the mutation are based on a tree structure.

The NSGA-II initially generates a population of random solutions, following the described coding. The size of this population ($pop_{size}$) is usually determined in a preliminary exploratory phase previous to the experimental optimization phase~\cite{GIBBS2015226}, as part of the parameters calibration. The population evolves along a limited number of generations ($gen_{num}$) by combining solutions to create an offspring. 

Solutions of the current generation are successively selected and combined, in pairs, to create individuals of the offspring. The selection operator is the binary tournament, defined as the default one for NSGA-II~\cite{deb2002fast}. The binary tournament selection chooses two solutions from the population at random, and the best of both solutions is finally selected for crossover. A second binary tournament selection is performed to select the second solution for the crossover. 

\begin{figure}[h!]
	\centering
	\includegraphics[width=\textwidth]{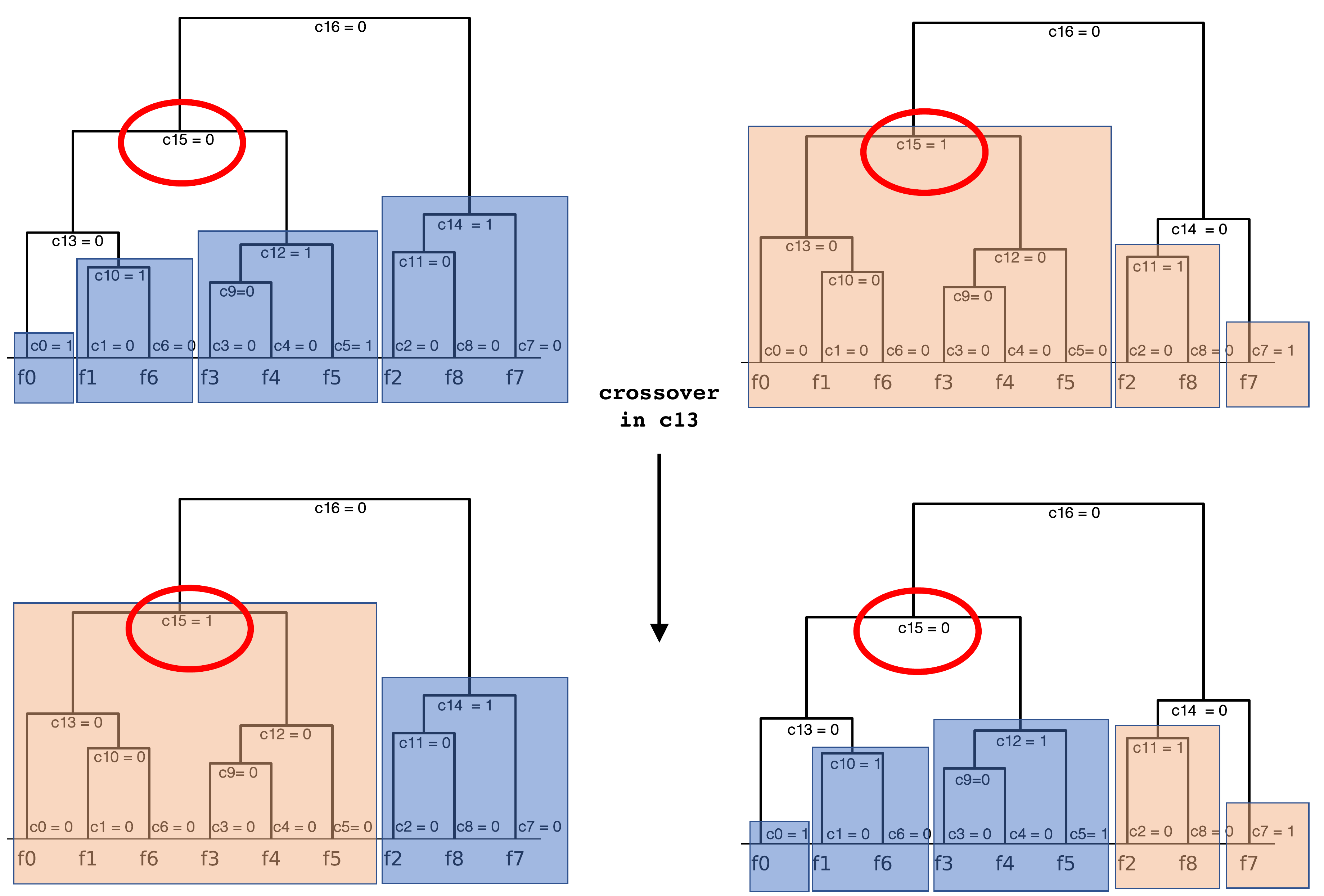}
	\caption{Example of the crossover operator, by randomly selecting candidate colony $C_{13}$.}\label{crossover}
\end{figure}

The two parent solutions are combined by applying the crossover operator, with a probability of $\rho_{cross}$. We use the sub-tree crossover~\cite{buijs2012genetic}, similar operator to the one-point crossover, but adapted for a tree-based structure instead of an array. The operator randomly chooses a node of the dendrogram, and the selected node and all its descendants (the sub-tree) are swapped between the two parents. Figure~\ref{crossover} shows and example.

Mutation operators are included in GAs to avoid falling into local minimums of the solution space. A mutation, with probability $\rho_{mut}$, is generated in the children obtained in the crossover by changing the values of some random positions of the chromosome. We define two mutation operators, colony-join and colony-split. They are randomly selected with probabilities $\rho_{mut}^{join}$ and $\rho_{mut}^{split}$  when a mutation is triggered.

\begin{figure}[h!]
	\centering
	\includegraphics[width=\textwidth]{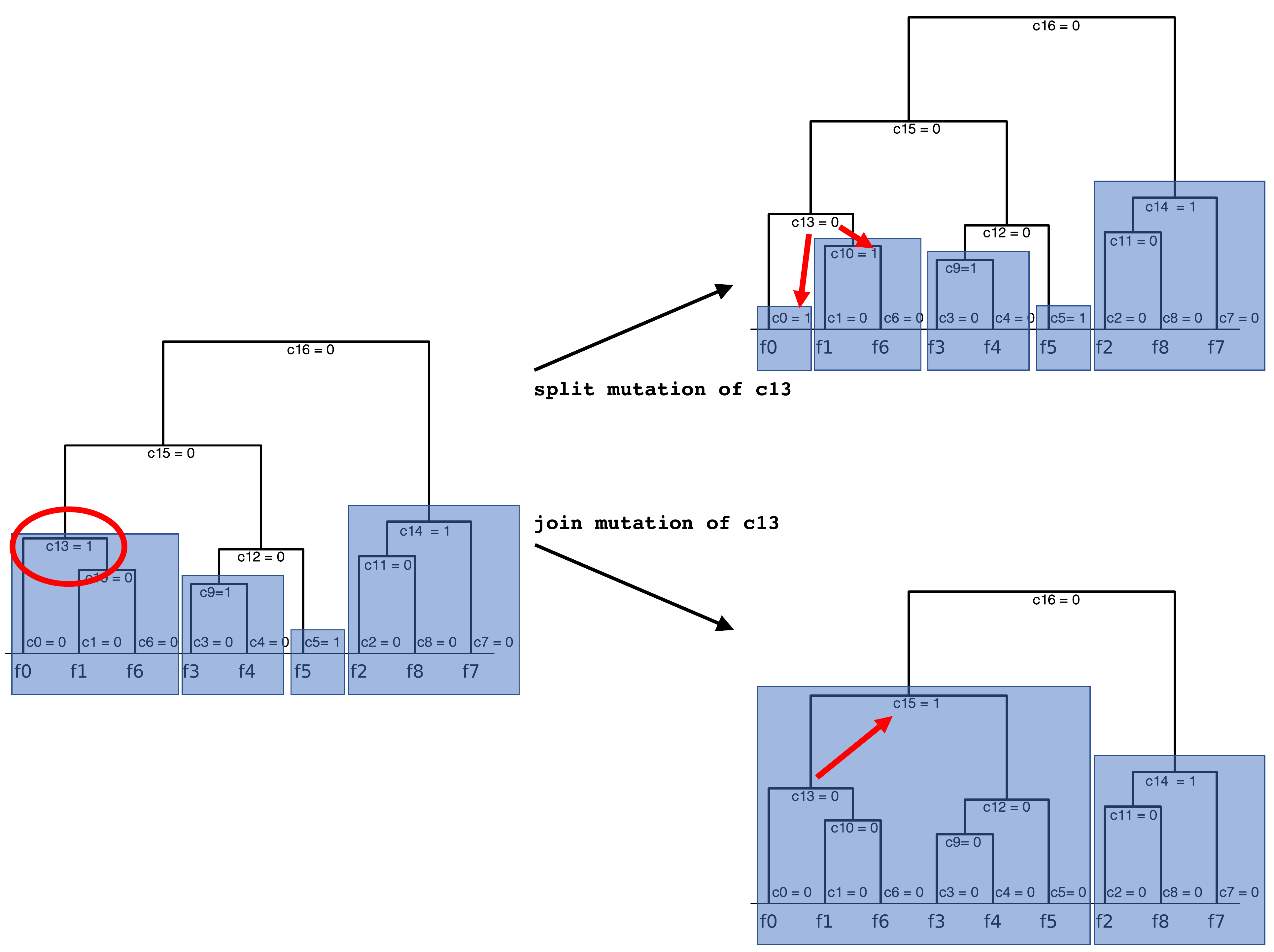}
	\caption{Example of the join and split colonies mutation, by randomly selecting candidate colony $C_{13}$.}\label{mutation}
\end{figure}

The colony-join mutation randomly chooses a current selected candidate colony (value 1 in the chromosome) and replaces this selected colony by its own parent. This action causes that the previous selected colony and its sibling (the other child of its parent) creates a new single colony. On the contrary, the colony-split mutation randomly chooses a current selected candidate colony (value 1 in the chromosome) and split this selected colony into the two colonies corresponding to its own children. Figure~\ref{mutation} shows and example of the result for both mutation operators.

Solutions obtained from a crossover or mutation can result into incorrect solutions due to the constraints defined in Section~\ref{sec_problemformulation}. We evaluate the solution suitability and if it is not correct, we transform it into a correct one using a repair operator. Some GA designs proposed to assign an infinite fitness to those incorrect solutions, but we rejected this alternative because our solutions are very likely to result in an incorrect one and, consequently, the population would include many incorrect solutions.

The repair operator checks whether: (a) a selected colony (content) is a subset of a bigger also selected colony (container); (b) all the devices are part of one colony. Two opposite repair operator are defined, the agglomerative one and the divisive, that are randomly selected to transform incorrect solutions into correct one, with probabilities $\rho^{agg}_{rep}$ $\rho^{div}_{rep}$ respectively. In the case of the agglomerative repair operator, it first keeps the container colony selected and deselects the content one. Non-included devices are grouped into colonies when possible, if not, one-element colony is created for the device. Both sub-tasks result on reducing the number of colonies. On the contrary, the divisive repair operator keeps the content colony selected and deselects the container one. A colony of one-element size is selected for each fog device that is not included in any colony. Both sub-tasks result on increasing the number of colonies. 

This process of the GA is repeated until the size of the offspring is the same than the population size, at witch point the new generation can be created. NSGA-II proposes to use elitism by default~\cite{NATESHA2021102972}. It means that only current generation's solutions dominated by offspring's solutions are replaced in the new generation. 

Apart from the repair operator, the fitness calculation is also particularly for our genetic implementation. The fitness calculation is composed of the estimated service response time ($response\_time$) and the placement algorithm execution time ($placement\_time$). The evaluation of both fitness components requires to perform the placement of the services in the fog devices of the colonies defined in each solution in the population. Thus, the placement of the services is determined by the real execution of a placement algorithm. We have selected a first-fit greedy algorithm for the service placement~\cite{lera2018availability}, but this algorithm could be easily replaced by another optimization heuristic~\cite{NAYERI2021103078}. For example, we also experimented with an Integer Linear Programming (ILP) approach, but it resulted intractable for colonies with a high number of devices. The details of our greedy implementation are explained in Section~\ref{sec_greedy}.

Accordingly, the greedy placement is executed for each solution in the population. The time of each of those executions is measured to obtain the $placement\_time$ component of the fitness. Once that the placement is performed, we can also evaluate the estimated application response time through Eq.~(\ref{eq_reponsetimeobjective}), obtaining the second component of the fitness ($response\_time$).

All the parameters that the previous steps defined as variables have a strong influence in the optimization process~\cite{birattari2009tuning}, and they need to be tuned previously to the experiment execution. The calibration of these parameters is usually defined in a pre-exploratory phase~\cite{10.1080/00207543.2013.784411}, where a range of values are studied to evaluate the most suitable parametrization. In our particular case, we studied the values of the population size, the number of generations and the probabilities of the different operators (Table~\ref{tab_geneticparameters}). 


\subsection{Greedy allocation of services}
\label{sec_greedy}

Once the colony layout is defined, for example for each solution in the genetic population, the placement of the services into the devices within the colony is required. Moreover, the fitness calculation of our GA requires this placement of the services.

Our proposed genetic optimization of the colony layout could be integrated with any placement algorithm for the services, but we have tested it with a simple but common proposal, a first-fit decreasing greedy algorithm~\cite{lera2018availability}. This type of algorithm orders the services to place in decreasing order using a given criterion, and each service is placed into the first device where it fits. The devices can be also ordered using a different criterion. If after considering all the devices in the colonies, the service is not placed anywhere, its placement is shifted to other neighboring colonies.

Algorithms~\ref{greedyplacement} shows the baseline of the placement algorithm. We base it in the idea that the most popular services (most requested) have to be placed in the closest devices to the users. Thus, the greedy algorithm initially lists all the services requested by the users in a colony. This list of services is ordered from the most to the least requested. The services are placed in the devices sequentially following this list order. The placement of the service of a iteration is determined by selecting the closest device to the users. To do that, all the devices in the colony are ordered by the distance between themselves and the users that request the service, from the closest to the furthest. If the service is requested from more than one edge node, we consider the mean value of the distances to all the nodes where the service is requested. The first device of the list where the service fits is selected for the placement. If there is not a device where the service fits, the placement of this service is shifted to the neighbouring colony. Accordingly, the greedy algorithm also requires to place services shifted from other colonies. This shifted services are also included in the initial ordered list of services, but giving priority to the services requested from the own colony. This iterative and greedy process of placing the services is repeated in parallel in each colony of the layout.

\begin{algorithm}[t]
	\caption{First-fit decreasing greedy service placement algorithm}
	\label{greedyplacement}
	\hspace*{\algorithmicindent} \textbf{Input} \\ \hspace*{\algorithmicindent*2} $requestedServices$ Services requested by users in the colony.\\
	\hspace*{\algorithmicindent*2} $colonyDevices$ Devices in the colony.\\
	\hspace*{\algorithmicindent*2} $shiftedServices$ Services with placements shifted to this colony
	\begin{algorithmic}[1]
		\Function{greedyPlacementInColony}{}
		\State $S_r \gets orderByPopularityDesc($requestedServices$)$
		\label{alg_rand_gen}
		\State $S_{sh} \gets orderByPopularityDesc($shiftedServices$)$
		\State $S_{ord} \gets S_r + S_{sh}$
		\For{$s\ $in$\ S_{ord}$}
		\State $userDevices \gets getDevicesWhereRequested(s)$
		\State $D \gets orderByMeanDistanceAsc(colonyDevices, userDevices)$
		\For{$d\ \in\ D$}
		\If {$fit(s,d)$} 
		\State $placement[s][d]=1$
		\State \textbf{break}
		\EndIf
		\EndFor \label{alg_end_for}
        \If {$\textbf{not}\ fit(s,d)$} 
		\State $c \gets getNeighbouringColony(this)$
		\State $c.shiftedServices.add(s)$
		\EndIf
		\EndFor
		\Return $placement$
		\EndFunction
	\end{algorithmic}
\end{algorithm}

\section{Experimental Evaluation}
\label{sect_experimentalevalutaion}

The experiments are designed to evaluate the optimization ratio of our proposal, to study the goodness of the results, and to compare our proposal with other two alternatives: only one colony that encompasses all the fog devices; and a human-assisted definition of colonies of a fixed size. But before addressing these issues, we first detail the experiment configuration in terms of the genetic optimization parameters and the experiment scenario outline.

As we explained in Section~\ref{sec_geneticoptimization}, the genetic parameter definition requires of a pre-exploratory calibration phase where different parameter values are tested to select the most suitable ones~\cite{10.1145/3377929.3398136}. In this preliminary phase, we finally set the parameters according the values in Table~\ref{tab_geneticparameters}. 

\begin{table}[t!]
	\caption{Genetic parameters values for the experimental phase.}
	\label{tab_geneticparameters}
	\centering
	\begin{tabular}{p{7.5cm}p{2.5cm}|r}
		\toprule
		\textbf{Parameter}
		&  & \textbf{Value}\\
		\midrule
		 Population size & $pop_{size}$ & 100 \\ \midrule
		 Number of generations & $gen_{num}$ & --- \\ \midrule
		 Probability of crossover & $\rho_{cross}$ & 0.95 \\ \midrule
		 Probability of mutation & $\rho_{mut}$ & 0.3 \\ \midrule
		 Probability to select between the mutation operators & $\rho_{mut}^{join}$, $\rho_{mut}^{split}$ & 0.5, 0.5 \\ \midrule
		 Probability to select between the repair operators & $\rho^{agg}_{rep}$ $\rho^{div}_{rep}$ & 0.5, 0.5  \\		
		\bottomrule
	\end{tabular}
\end{table}	

The experiment scenario is defined by the fog infrastructure, the applications deployed in the system and the users connected to it. The configuration parameters are summarized in Table~\label{tab_experimentscenarioparameters}. We chose the Barabasi-Albert topology for the network infrastructure and the Betweenness centrality index for the coordinator selection because they resulted on the topology and the centrality index that most benefit the network times in a fog colony-based layout~\cite{guerrero2018influence}. The system workload were defined in terms of the number of gateways and the number of users per gateway that request each application (application popularity). The rest of parameters were based on the experiments in Gupta et al.~\cite{ifogsimgupta17}. 

\begin{table}[t!]
	\caption{Scenario configuration values for the experimental phase.}
	\label{tab_experimentscenarioparameters}
	\centering
	\begin{tabular}{p{7.5cm}|r}
		\toprule
		\textbf{Feature}
		&   \textbf{Value}\\
		\midrule
\multicolumn{2}{l}{\textbf{Infrastructure}}  \\ \midrule
		 Number of fog devices & 100..300 \\ \midrule
		 Infrastructure topology & Barabasi-Albert \\ \midrule
		 Coordinator selection & Betweenness centrality\\ \midrule
		 Fog network latency (ms) & 2--6 \\ \midrule
		 Cloud network latency (ms) & 100 \\ \midrule
		 Fog device resources (resource units) & 1--4 \\ \midrule
         Percentage of gateways fog devices (\%) & 25 \\ \midrule
\multicolumn{2}{l}{\textbf{Application}}  \\ \midrule
		 Number of applications & 20..60 \\ \midrule
		 Application resources (resource units) & 1--2 \\ \midrule
		 
\multicolumn{2}{l}{\textbf{User}}  \\ \midrule
		 User inter-request time (ms) & 5--10 \\ \midrule
         Application popularity per gateway (\%) & 0--75  \\		
		\bottomrule
	\end{tabular}
\end{table}

The experiments are executed using our own implementation of the NSGA-II algorithm in Python 3~\footnote{Source code available on \url{https://github.com/acsicuib/GAFogColonies}}. Additionally, we have also implemented two other optimization proposals to compare the results with, the one with a fixed size of colonies and the one with only one colony. 

The first control algorithm does not consider colonies in the fog layout, and it is functionally equal to say that there is only one colony that includes all the devices. Consequently, it is not necessary a colony optimization process for this case. The optimization process is limited to a global optimization of the service placement. We have used the same service placement algorithm that in our proposal for a fair comparison of the influence of the optimization of the colony layout. We label the results of this approach as \textit{one-colony}. 

Accordingly to the related research included in Section~\ref{sect_relatedwork}, we have also included a second control algorithm that homogeneously defines the fog colonies using a fixed colony size. The size of the colony is pre-determined in a human-assisted process~\cite{guerrero2018influence}. In this previous phase, we have determined an average colony size of 5 devices as the best solution when an homogeneous colony size is used. We label the results of this approach as \textit{fixed-size}.

\subsection{Results}

The experiments are varied in terms of the number of nodes (100, 200, and 300) and the number of applications (20, 40, and 60), resulting in the execution of 9 experiment scenarios for each of the three colony definition algorithms (GA, \textit{one-colony}, and \textit{fixed-size}). Each of these experiment scenarios is labeled as \textit{XnodesYapps}, where the X is the number of nodes and Y is the number of apps of the experiment.

\begin{figure} [t!]
\centering
\includegraphics[trim=65 5 80 15,clip,width=0.75\textwidth]{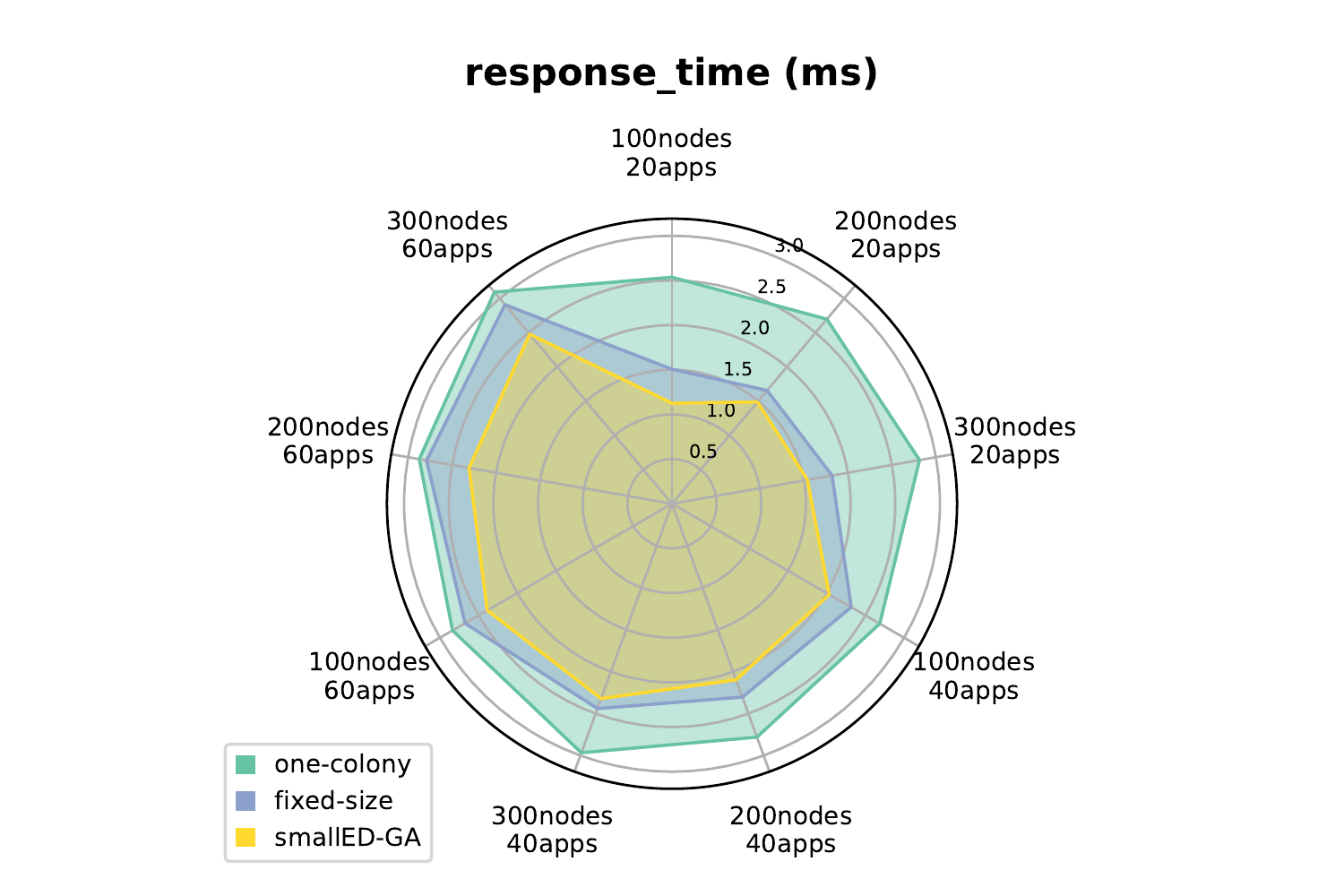}
\caption{Comparison of the $response\_time$ for the GA selected solution and the other two control algorithms.}
\label{fig_response}
\end{figure}




\begin{figure}[t!]
\centering

\subfloat[Results for the GA selected solution and the other two control algorithms. A zoom of the data results is shown in Figure~\ref{fig_placement2}\label{fig_placement3}]{\includegraphics[trim=65 5 80 15,clip,width=0.6\textwidth]{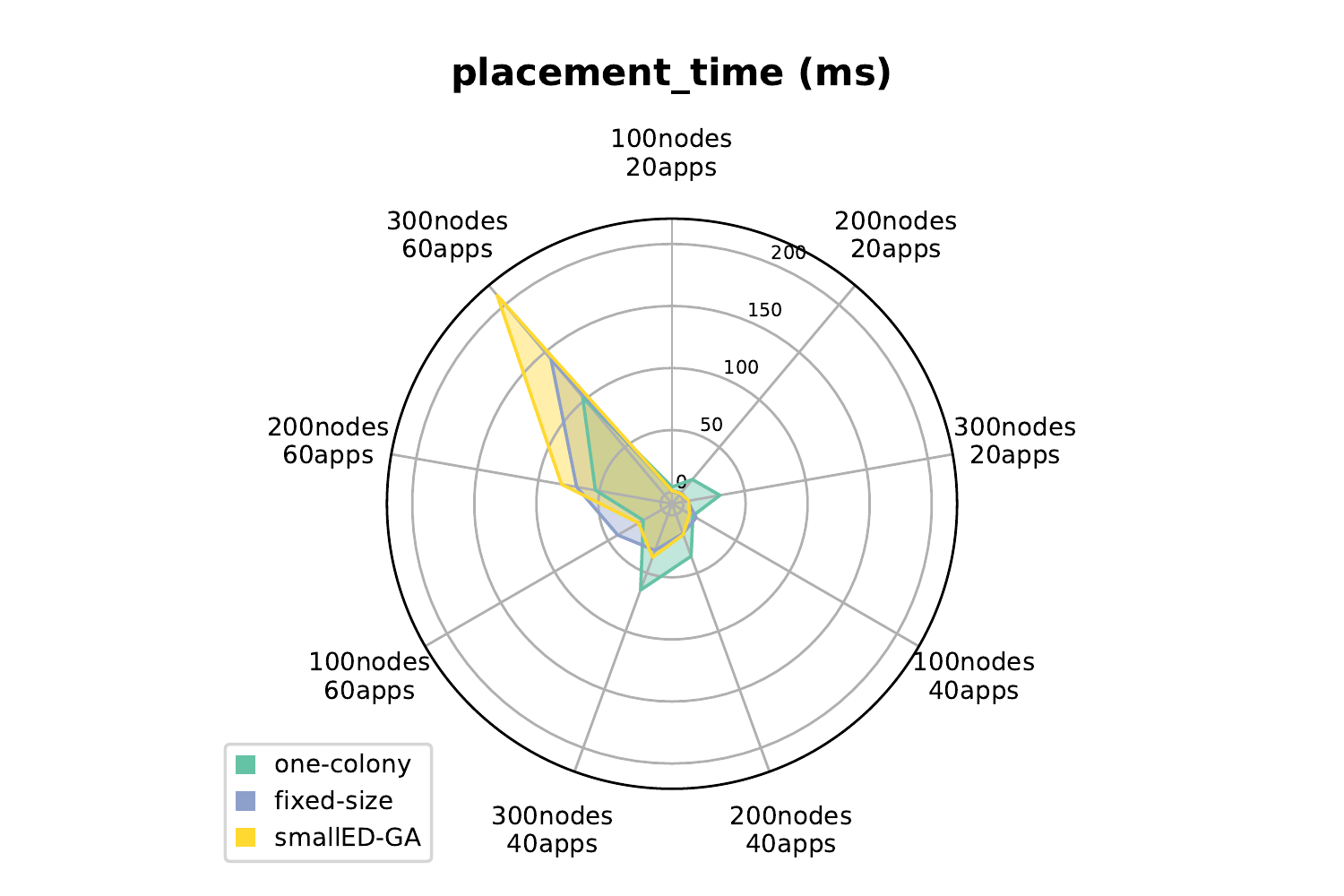}}
\\

\subfloat[Zoom of the results of Figure~\ref{fig_placement3}\label{fig_placement2}]{\includegraphics[trim=65 5 80 15,clip,width=0.6\textwidth]{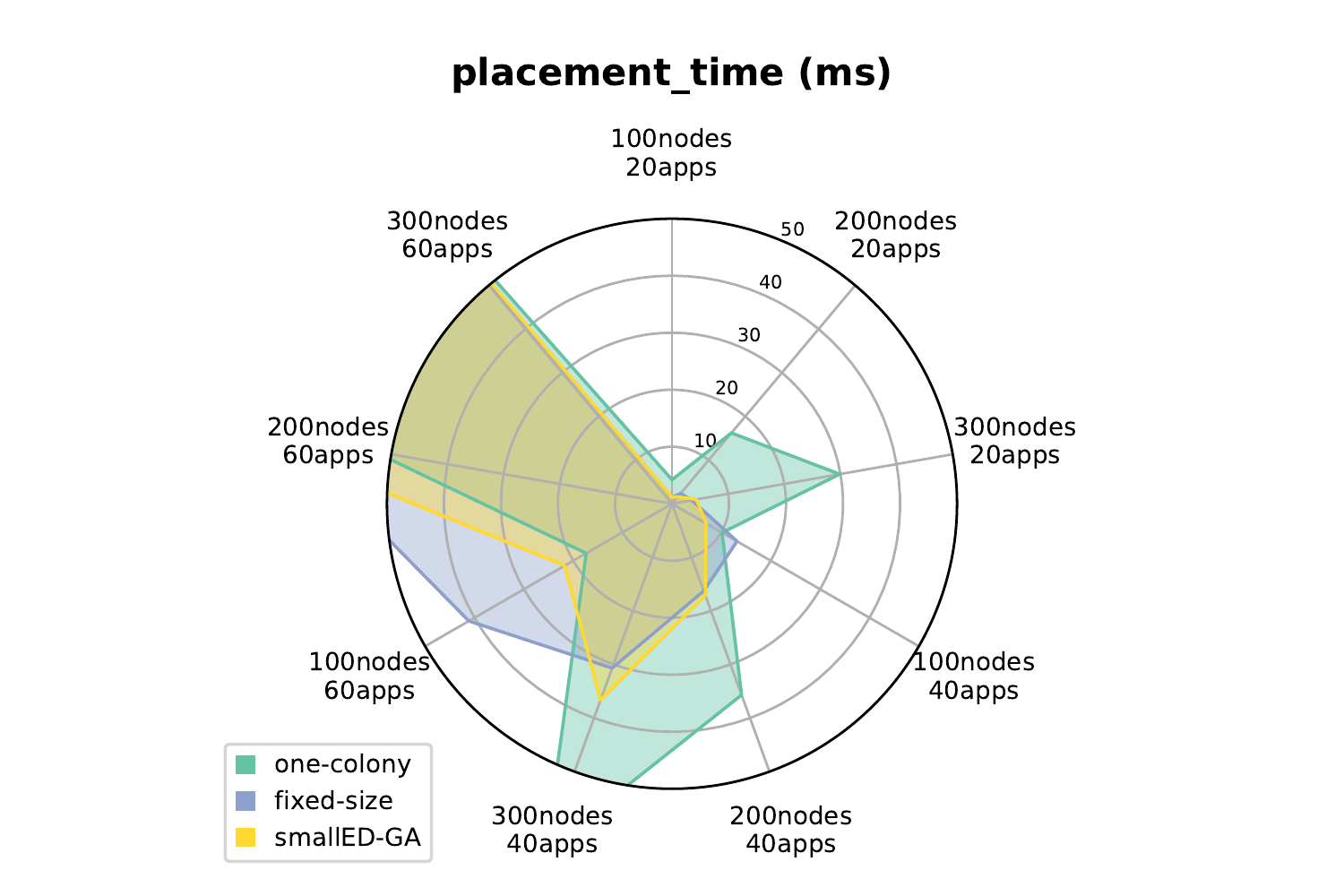}}

\caption{Comparison of the $placement\_time$.}
\label{fig_placement}
\end{figure}


Remember that the NSGA-II returns a set of solutions, contrary to the two control algorithms, which return one solution. We first compare the \textit{fixed-size} and \textit{one-colony} solutions with one solution selected from the Pareto set. We select the solution with the smallest Euclidean distance between the normalized values of the solution's objectives and the origin point of the objective space. We label this solution as \textit{smallED-GA} Figures~\ref{fig_response} and~\ref{fig_placement} respectively  compare the $response\_time$ and the $placement\_time$ for the two control algorithms and the \textit{smallED-GA} solution selected from the Pareto set of the GA. For a clearer comparison of the results, the plot is zoomed in for the case of $placement\_time$ in Figure~\ref{fig_placement2}.



For the case of the $response\_time$, the solution selected from the GA is better, because it shows the smallest value for the average response time of the applications, in all the 9 experiment scenarios. On the contrary, the \textit{one-colony} is the worst of the three cases. This is explained because the GA is the algorithm that creates a fog colony layout that better balances the size of the colonies.  

On the contrary, the $placement\_time$ does not show a clear trend, and the selected GA solution is only better in two of the 9 experiment scenarios. This might lead us to conclude that the GA shows a bad behaviour in terms of the $placement\_time$. But it is important to remember that  we are considering only one solution from the Pareto set of the GA, and the process of selection of this solution has a strong influence. A deeper comparison, considering all the Pareto set is required, and we have graphically represented the objective space of the Pareto front to perform this deeper analysis.

\begin{landscape}

\begin{figure}[h!]
	\centering
	\begin{raggedleft} 
	\scalebox{0.47}{\includegraphics[trim=15 5 35 15,clip]{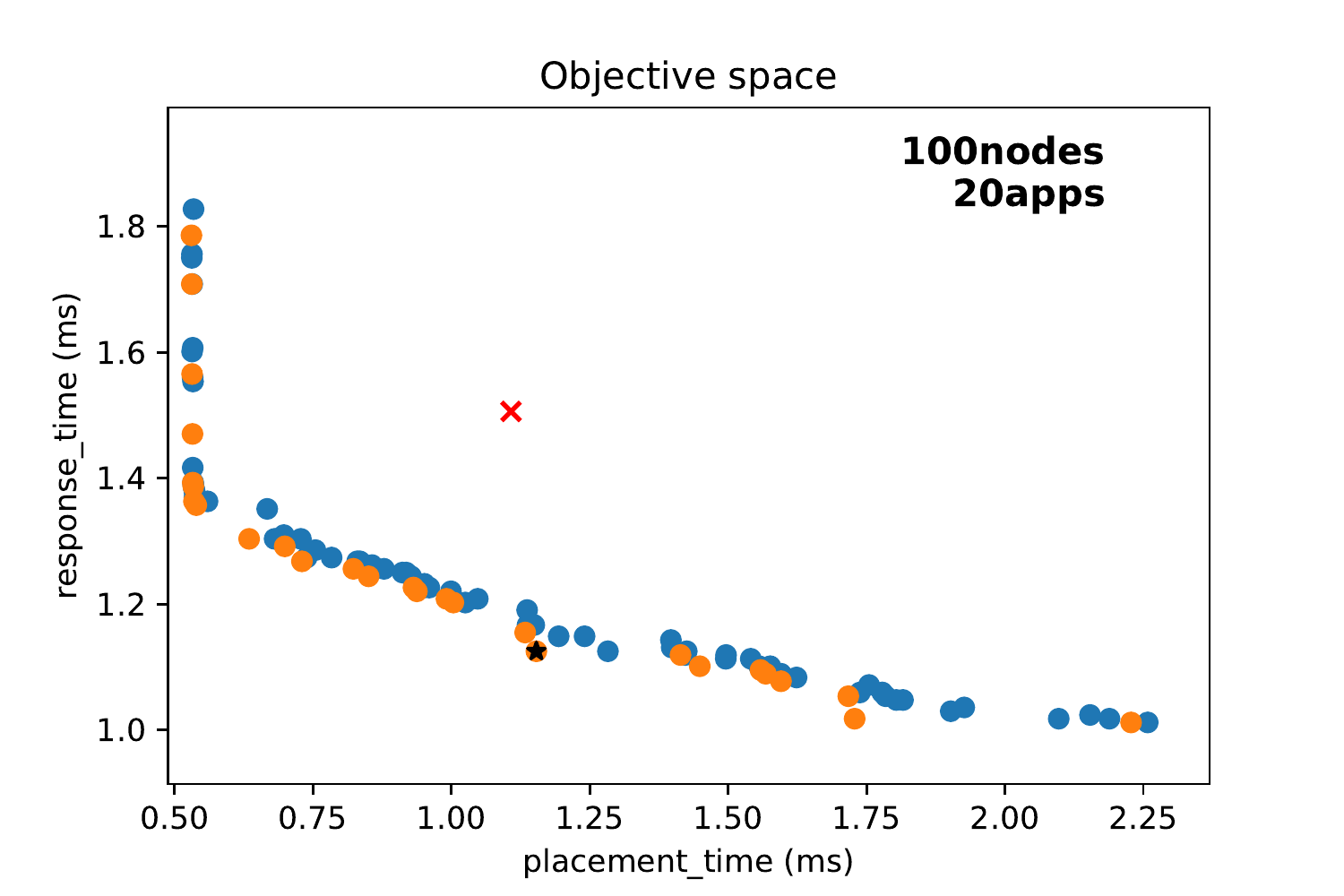}}\hfill
	\scalebox{0.47}{\includegraphics[trim=15 5 35 15,clip]{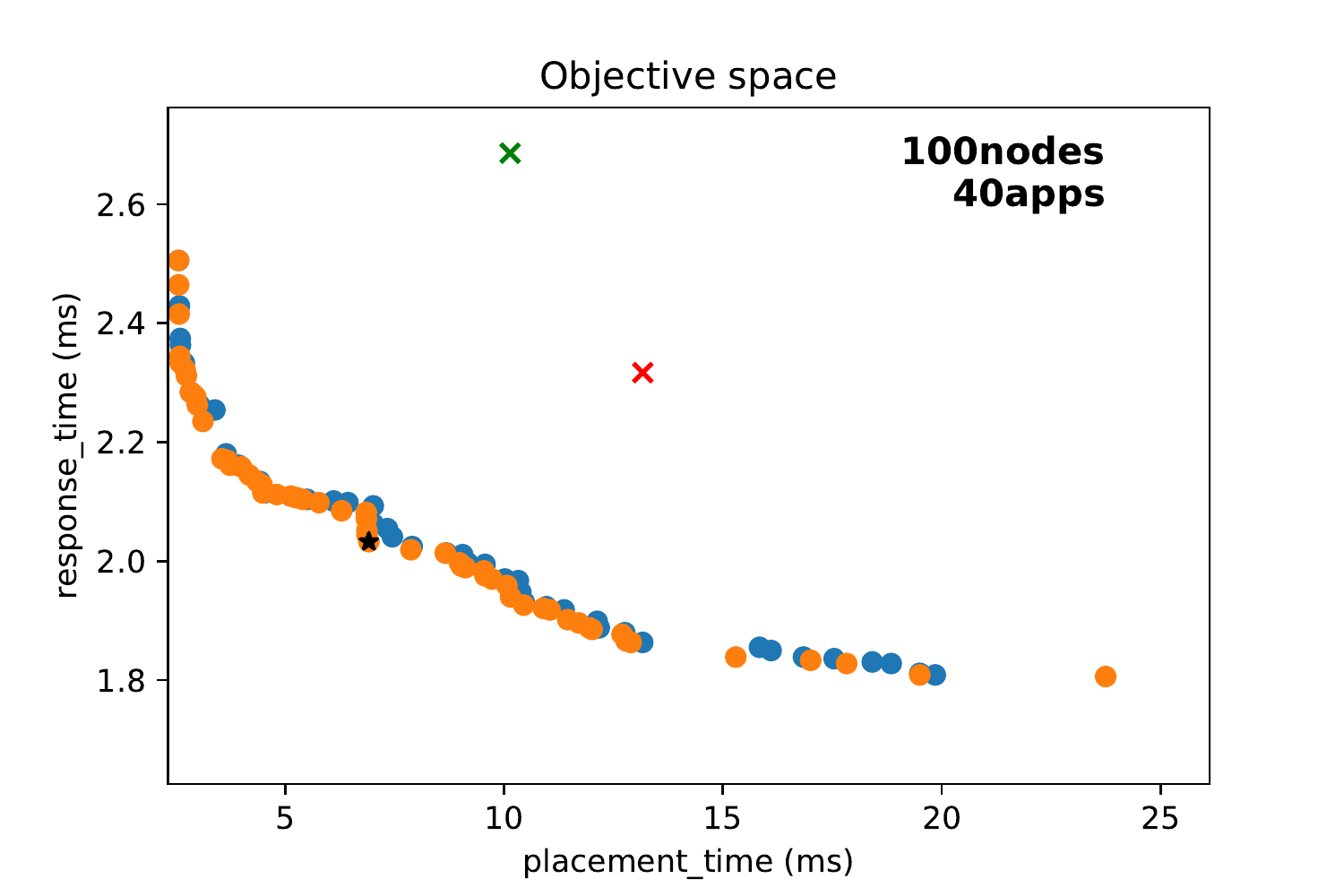}}\hfill
	\scalebox{0.47}{\includegraphics[trim=15 5 35 15,clip]{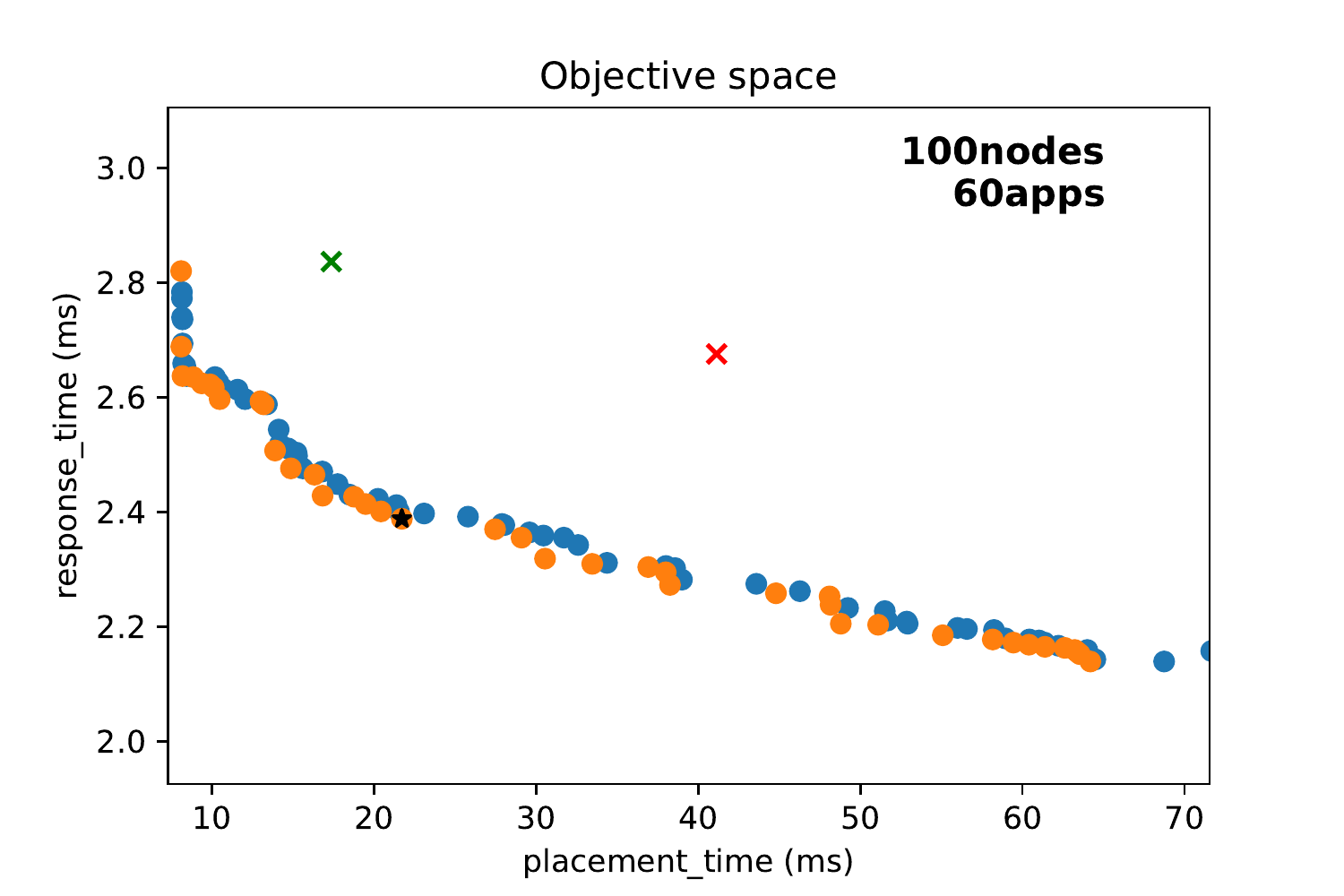}}
	\end{raggedleft}
	\begin{raggedleft} 
	\scalebox{0.47}{\includegraphics[trim=15 5 35 15,clip]{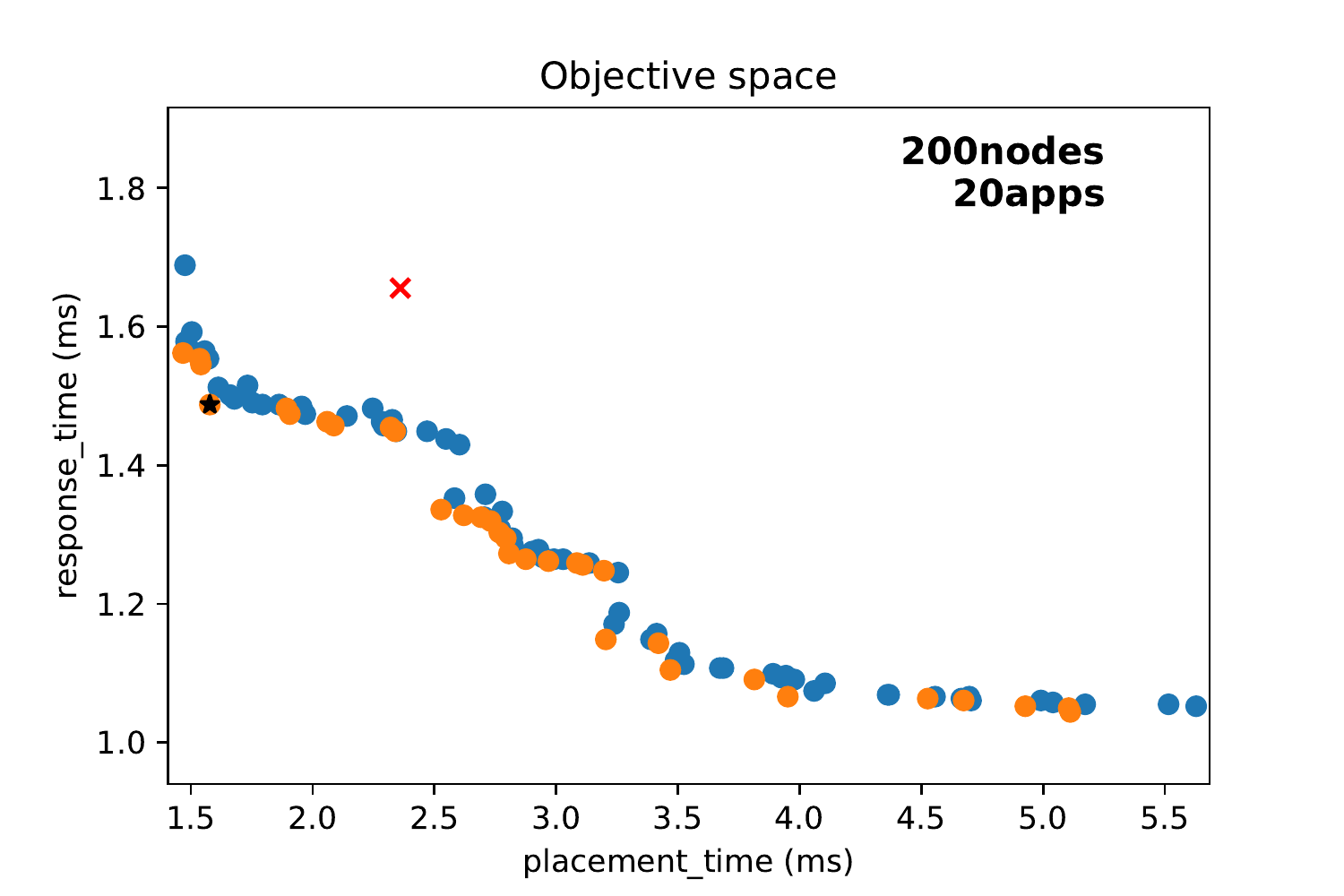}}\hfill
	\scalebox{0.47}{\includegraphics[trim=15 5 35 15,clip]{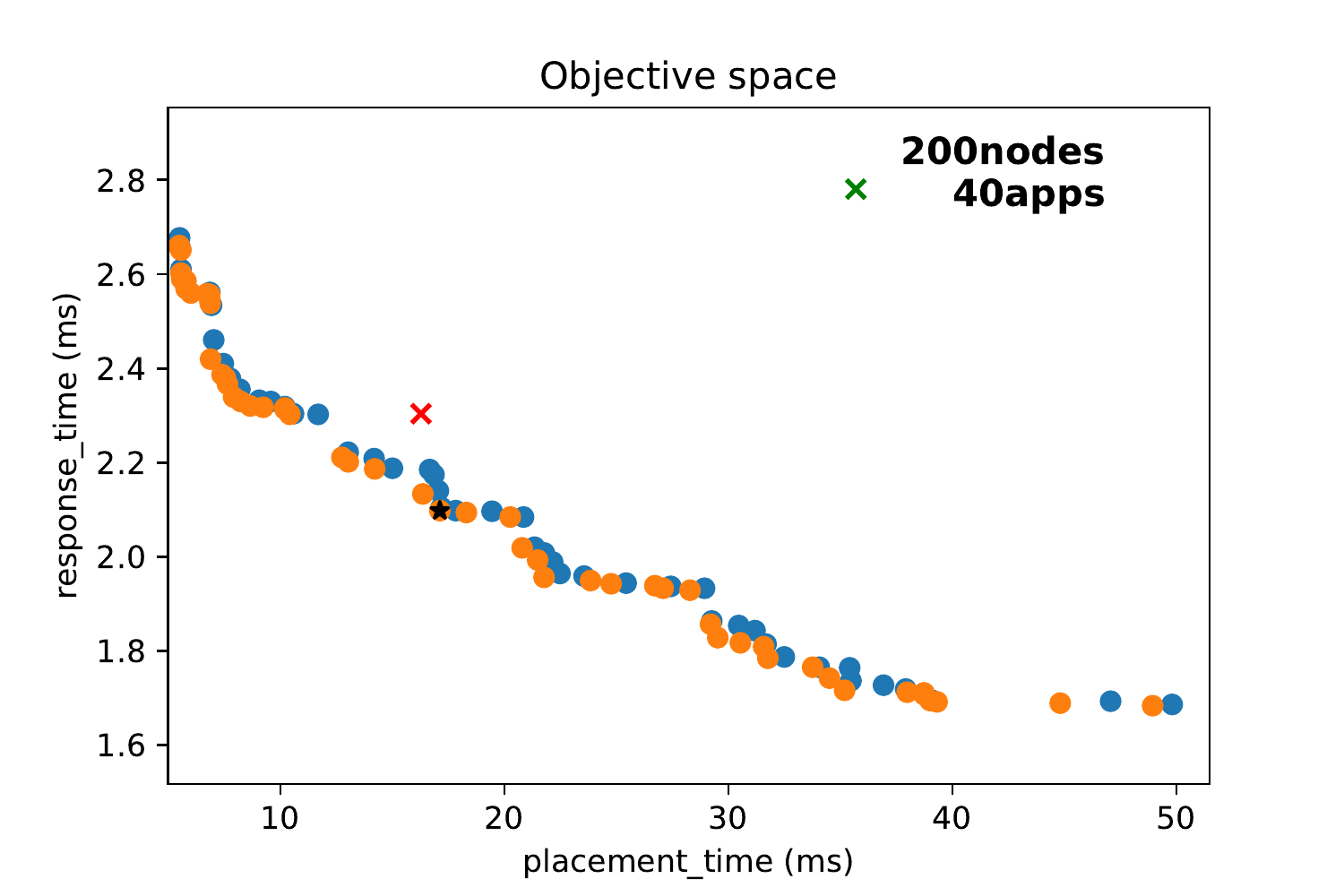}}\hfill
	\scalebox{0.47}{\includegraphics[trim=15 5 35 15,clip]{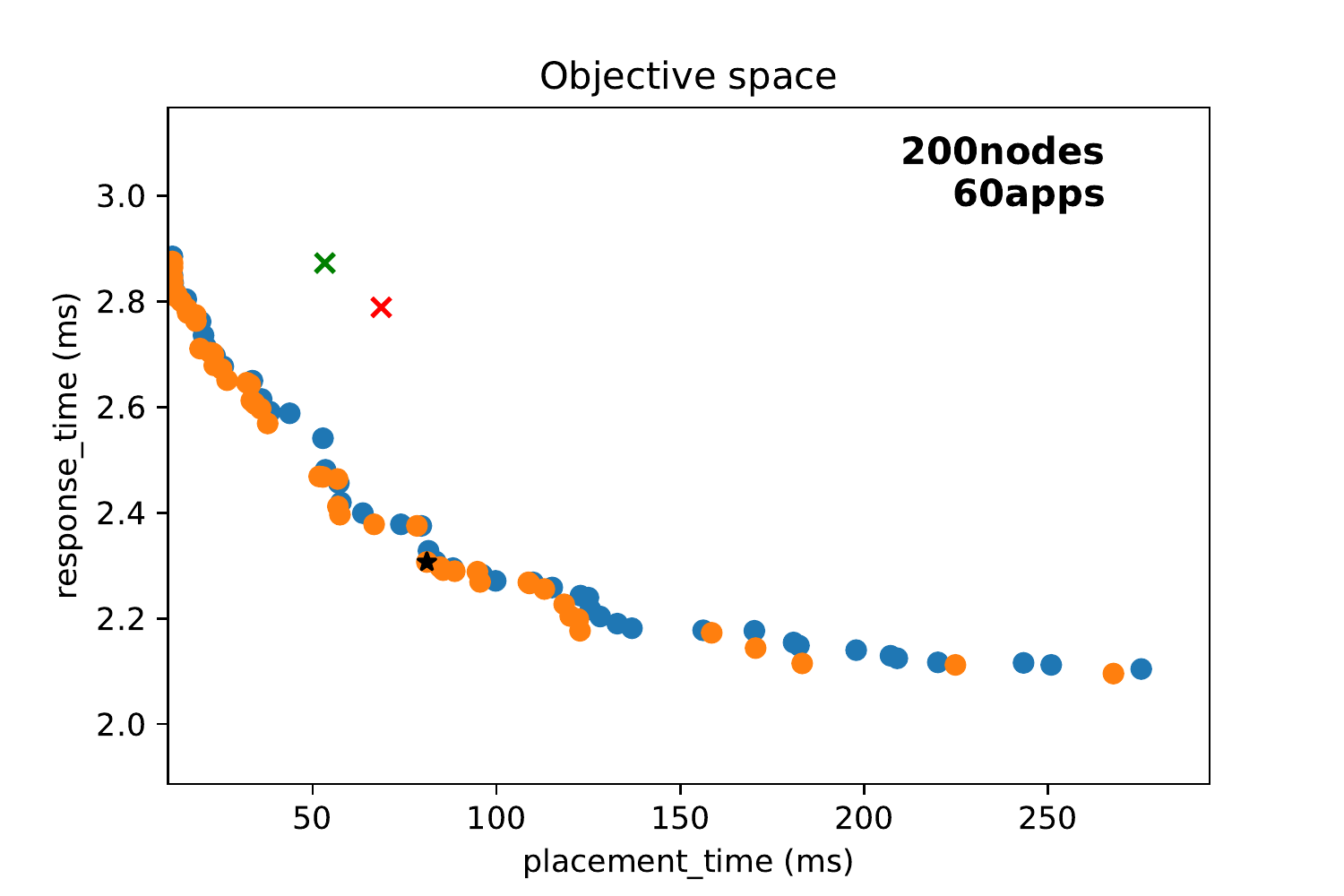}}
	\end{raggedleft}
	\begin{raggedleft} 
	\scalebox{0.47}{\includegraphics[trim=15 5 35 15,clip]{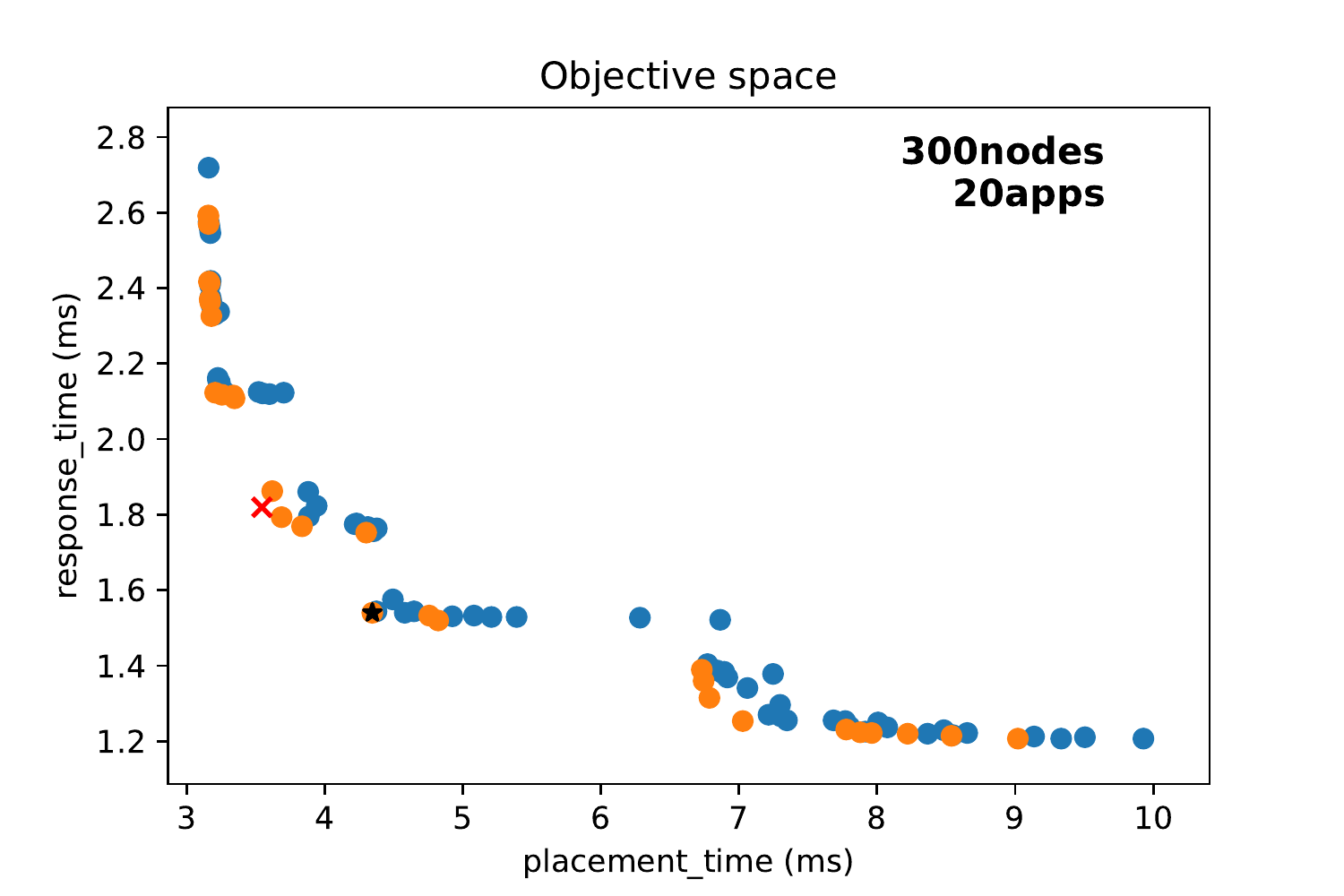}}\hfill
	\scalebox{0.47}{\includegraphics[trim=15 5 35 15,clip]{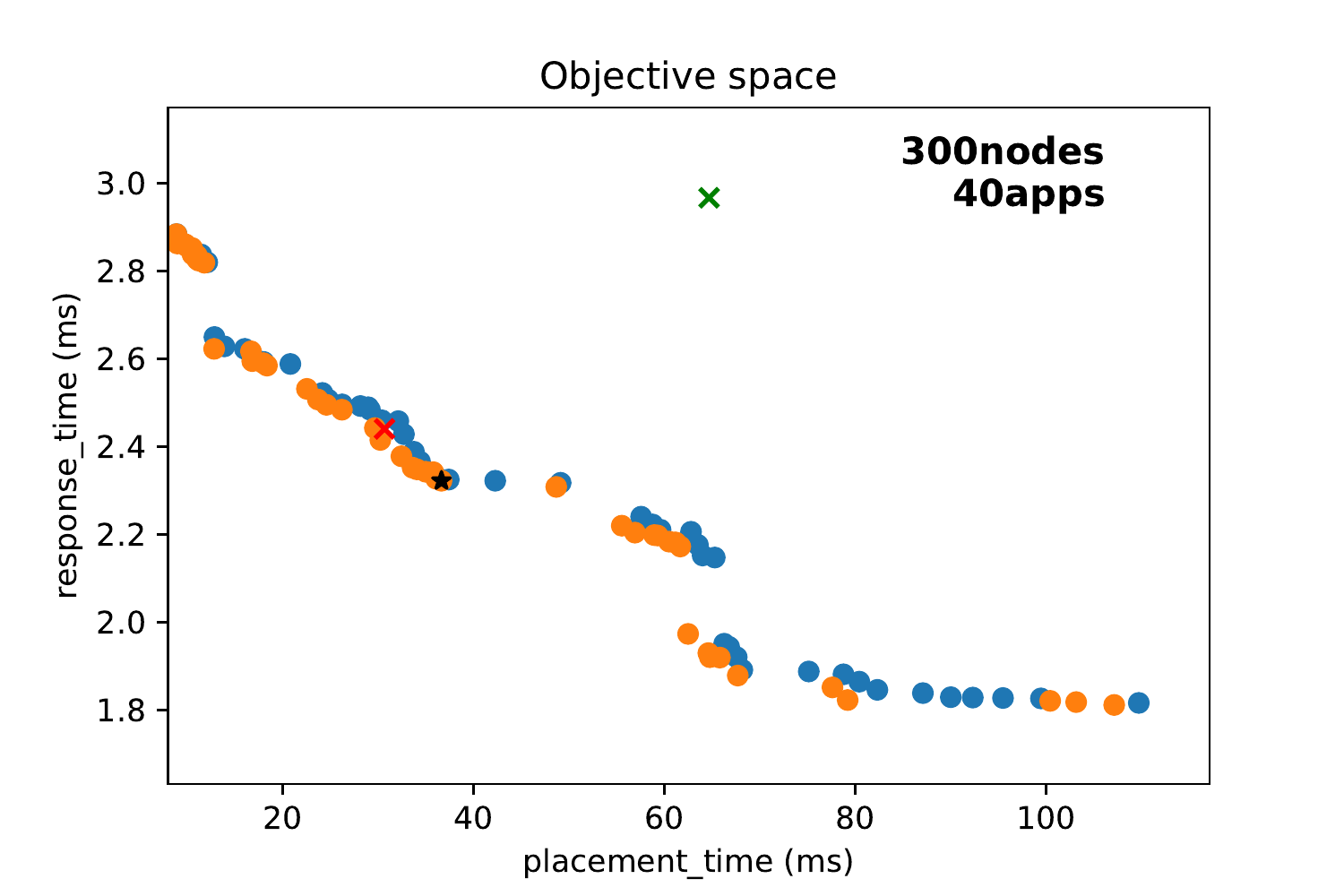}}\hfill
	\scalebox{0.47}{\includegraphics[trim=15 5 35 15,clip]{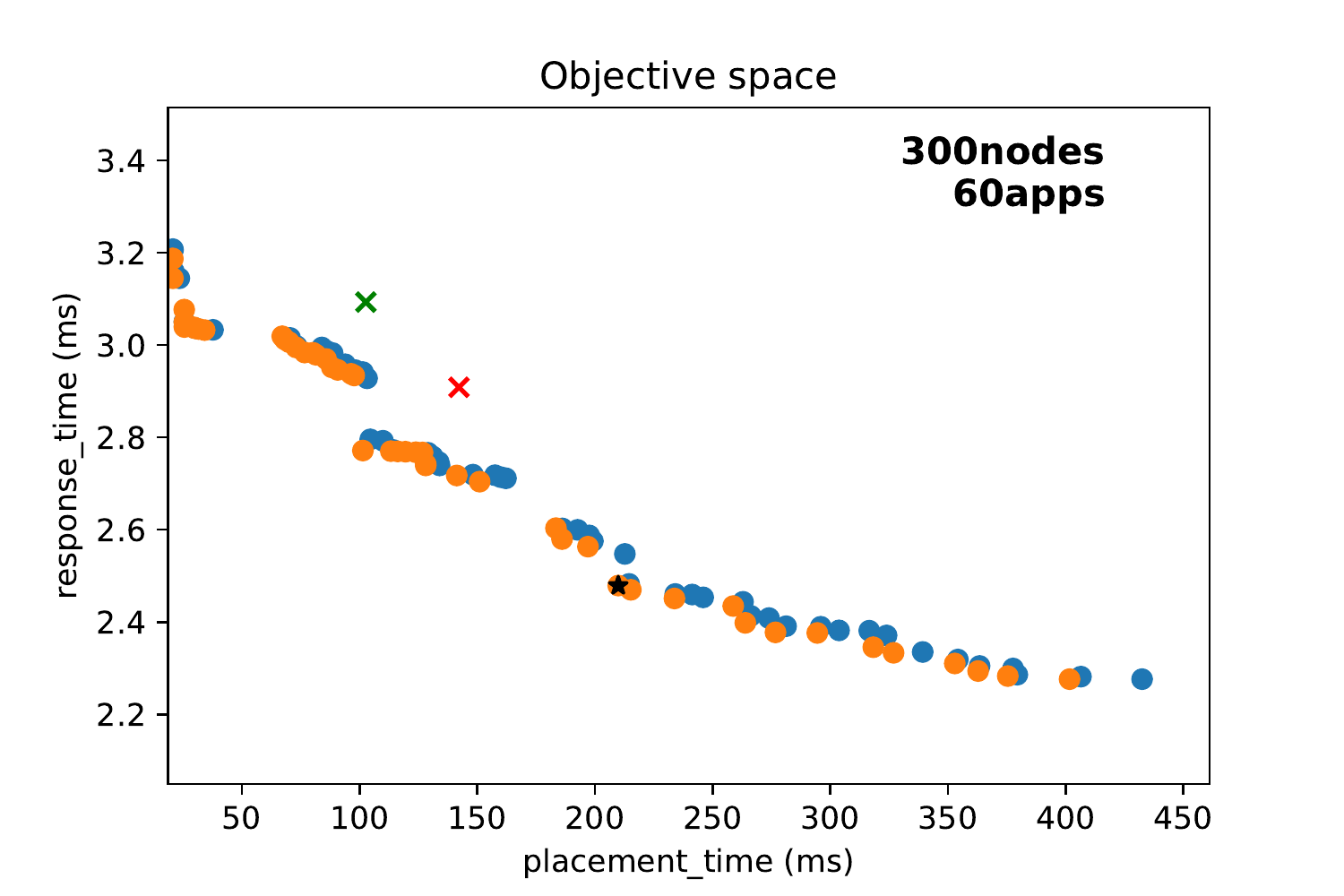}}
	\end{raggedleft}
	\includegraphics{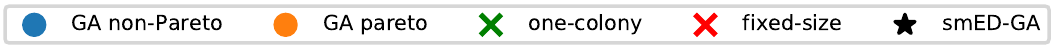}
	\caption{Scatter plots that represent the objective space of the experiments.}\label{fig_objectivespaces}
\end{figure}

\end{landscape}

Figure~\ref{fig_objectivespaces} represents the objective space of the last generation for each of the 9 experiments, including the objective values for the solutions in the Pareto front (orange coloured points), values for solutions in other fronts (blue colored points), the solution selected with the Euclidean distance (marked with a black star), the objective values of the solution obtained with the \textit{fixed-size} algorithm (red colored cross), and with the \textit{one-colony} algorithm (green colored cross).

Figure~\ref{fig_objectivespaces} shows that the Pareto fronts of the GA dominate the solutions of the control algorithms in all the experiment scenarios except one, the experiment with 300 nodes and 20 apps. If the Pareto front dominates the other two control solutions, we are able to find at least one solution in the Pareto set that reduces both optimization objectives. This confirm us that the bad results obtained in Figure~\ref{fig_placement} are due to the solution selection procedure and not related with the GA optimization. The selection of the solution through the smallest Euclidean distance clearly benefit the $response\_time$ and damaged the $placement\_time$.

This highlights the idea that having a Pareto set of solutions offers a wider number of alternatives to apply specific criteria for the selection of the solution that better fits each specific case. Accordingly, the quality of the Pareto front is related to the distribution of the solution along the objective space and the maximization of the range of the values of each objective~\cite{knowles2002metrics}. This is an inherent advantage of multi-objective optimization techniques that are based on dominance and that result on a Pareto set.

In relation to the experiment with 300 nodes and 20 apps, the GA was not able to dominate the solution obtained in \textit{fixed-size}, but the differences in the objective values were very small. After a deeper analysis of this case, we confirmed that, with a higher number of iterations of the GA, the Pareto front improved the solution obtained with the \textit{fixed-size}. Concretely, the Pareto front in generation 137 was the first one that dominated the \textit{fixed-size} solution. Consequently, the lower optimization of the GA, in this experiment scenario, was due to an underestimated calibration of the GA parameters. To evaluate this calibration, it would be interesting to evaluate the first generation in which the Pareto front dominates the two control solutions.

\begin{table}[t!]
	\caption{Comparative metrics for the non-dominated solution sets.}
	\label{tab_paretocomparative}
	\centering
	\begin{tabular}{l|r|r}
		\toprule
		\textbf{Experiment} & $\mathcal{S}(GA)$ & Generation when \\  &&$\mathcal{C}(\text{GA},\text{one-colony}) = $  \\
		&&$= \mathcal{C}(\text{GA},\text{fixed-size}) = 1 $  \\
        \midrule
100nodes
20apps  &  0.772  &  10  \\ \midrule
200nodes
20apps  &  1.085  &  31  \\ \midrule
300nodes
20apps  &  1.241  &  137  \\ \midrule
100nodes
40apps  &  1.036  &  1  \\ \midrule
200nodes
40apps  &  1.087  &  24  \\ \midrule
300nodes
40apps  &  1.072  &  38  \\ \midrule
100nodes
60apps  &  1.080  &  1  \\ \midrule
200nodes
60apps  &  1.038  &  6  \\ \midrule
300nodes
60apps  &  1.036  &  77  \\		
		\bottomrule
	\end{tabular}
\end{table}

The size of the objective space covered by the non-dominated solutions of the GA and the coverage are common metrics to perform deeper analysis of the Pareto fronts and deeper comparison with the control experiments. The size of the objective space covered by the non-dominated solutions of the GA~\cite{zitzler1999multiobjective}, $\mathcal{S}(GA)$, is a quality unary indicator which measures the distribution and the range of the objective values covered by the Pareto front. The bigger the volume of the objective space, the better the Pareto front is.  Secondly, the coverage\footnote{The coverage $\mathcal{C}(A,B)$ is the fraction of solutions in B that are dominated by the solutions in A.} of the control algorithm solutions in regard with the Pareto front of the GA~\cite{zitzler1999multiobjective}, $\mathcal{C}(\text{one-colony},\text{GA})$ and $\mathcal{C}(\text{fixed-size},\text{GA})$, gives us an idea of the alternative solutions to the control ones that are obtained with the GA. The smaller the value of the coverage, the better the solutions of the GA are.

After analyzing the Pareto fronts in Figure~\ref{fig_objectivespaces}, we can easily conclude that  $\mathcal{C}(\text{one-colony},\text{GA})\ =\   \mathcal{C}(\text{fixed-size},\text{GA})\ =\ 0.0$ and  $\mathcal{C}(\text{GA},\text{one-colony})\ =\ \mathcal{C}(\text{GA},\text{fixed-size}) = 1.0$ in generation number 100 for all the experiment scenarios, except in the case of 300 nodes and 20 apps that achieved these values in generation number 137. We consider that, instead of analyzing the final values of the coverages, it is more interesting to calculate the first generation in which the Pareto front dominates the two control solutions, i.e.,  $\mathcal{C}(\text{GA},\text{one-colony})\ =\ \mathcal{C}(\text{GA},\text{fixed-size}) = 1.0$. Accordingly, Table~\ref{tab_paretocomparative} shows the number of this first generation and the size of the objective space covered by the non-dominated solutions of the GA.

\begin{figure}[h!]
	\centering
	\begin{raggedleft} 
	\scalebox{0.45}{\includegraphics[trim=15 5 35 15,clip]{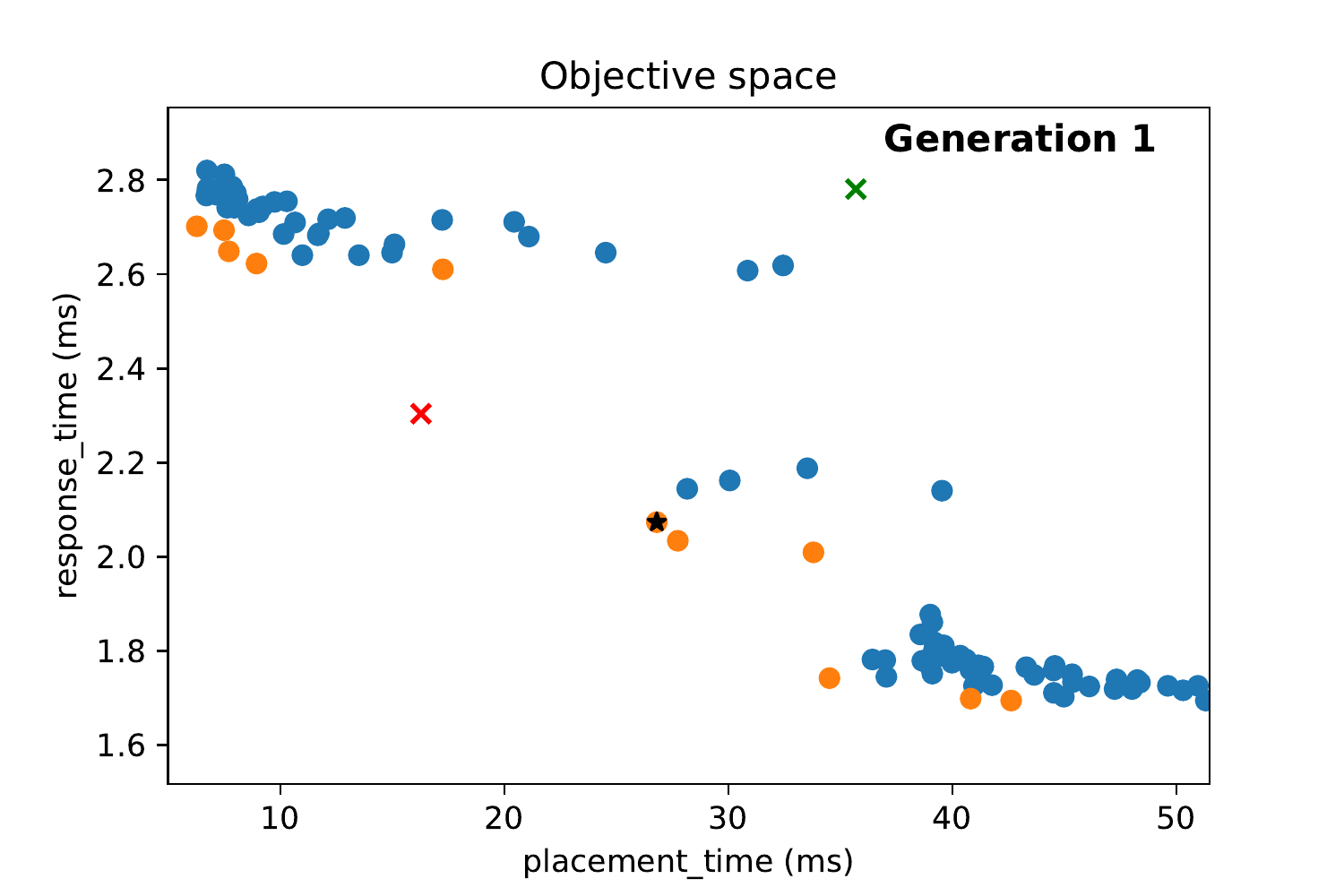}}\hfill
	\scalebox{0.45}{\includegraphics[trim=15 5 35 15,clip]{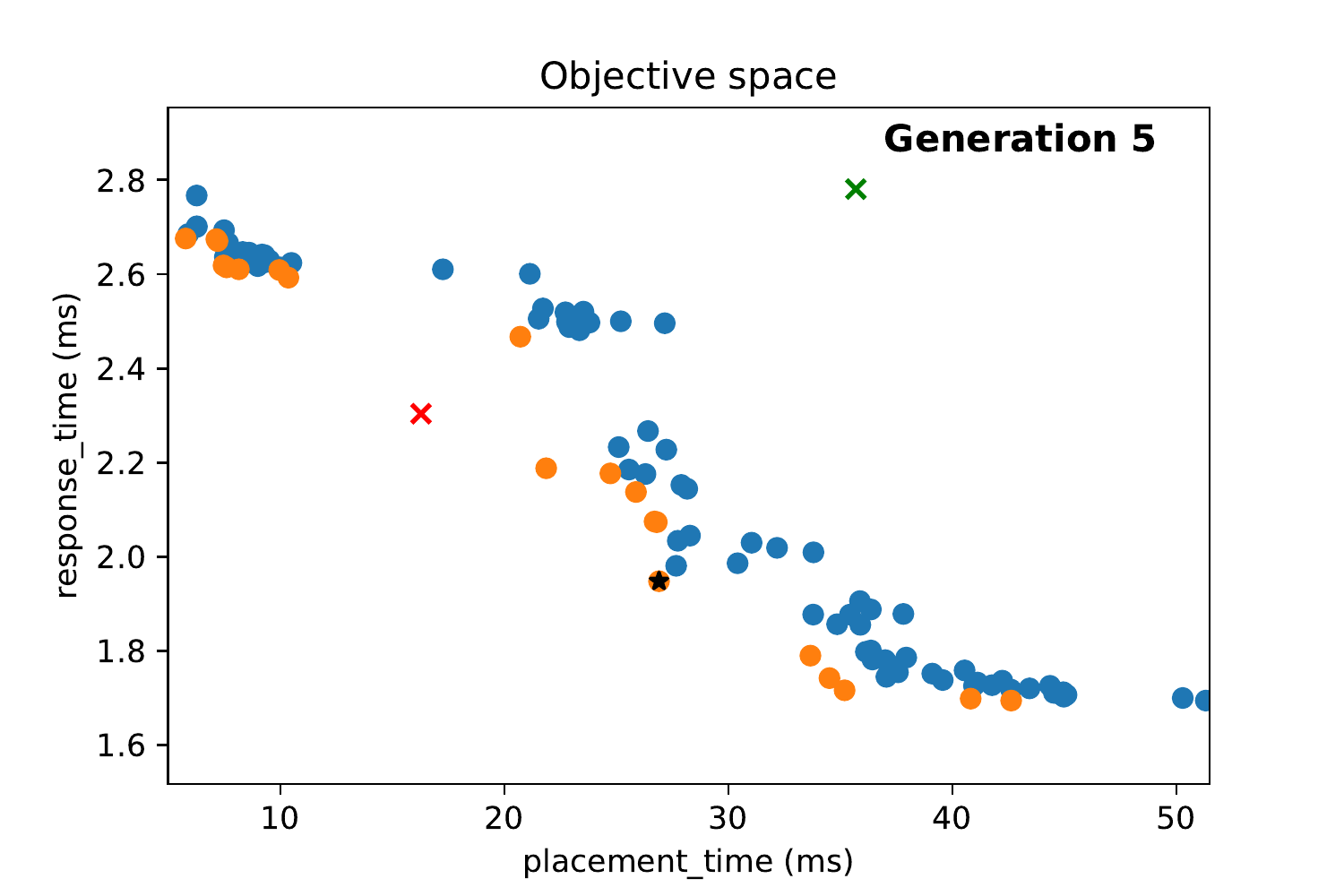}}
	\end{raggedleft}
	\begin{raggedleft} 
	\scalebox{0.45}{\includegraphics[trim=15 5 35 15,clip]{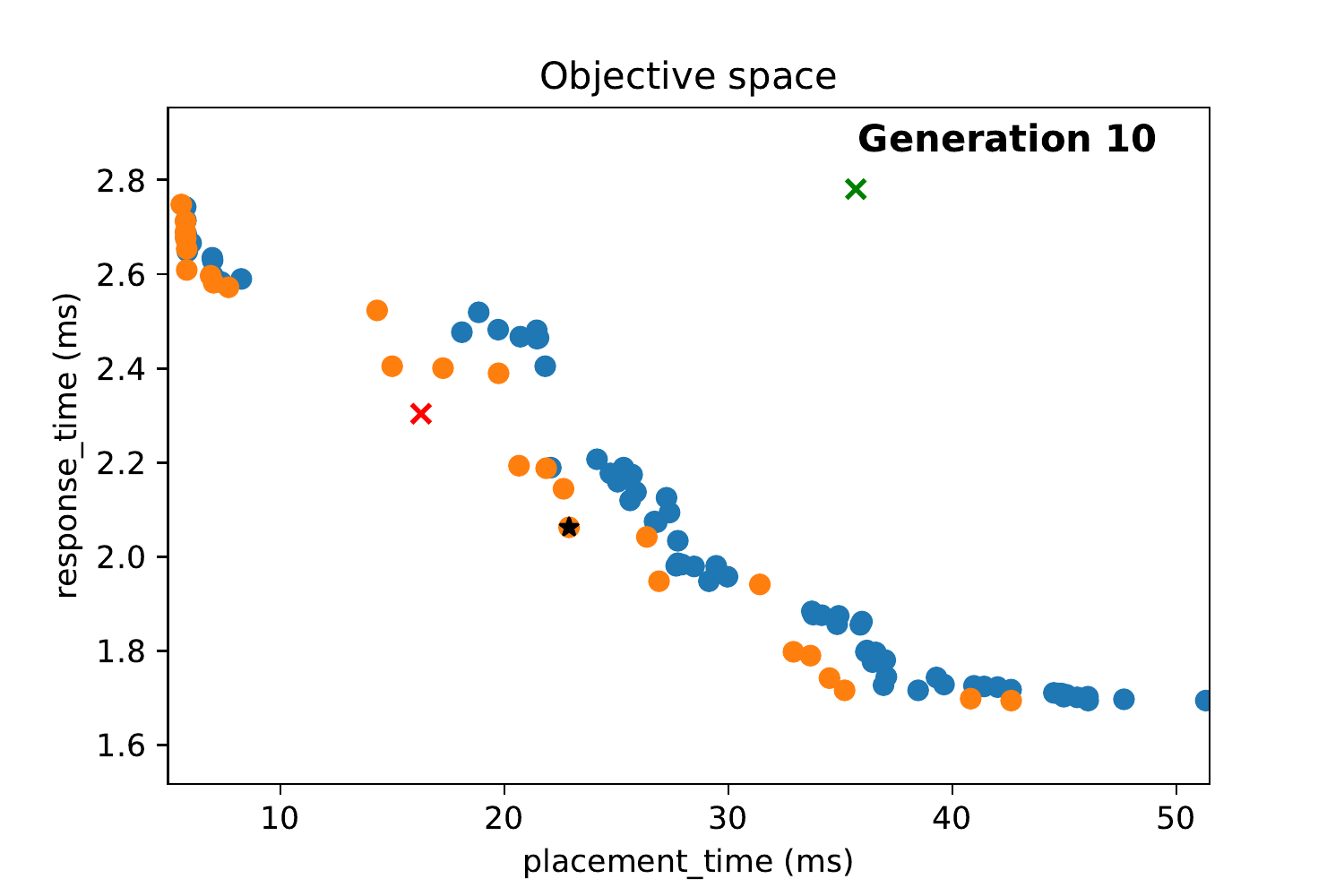}}\hfill
	\scalebox{0.45}{\includegraphics[trim=15 5 35 15,clip]{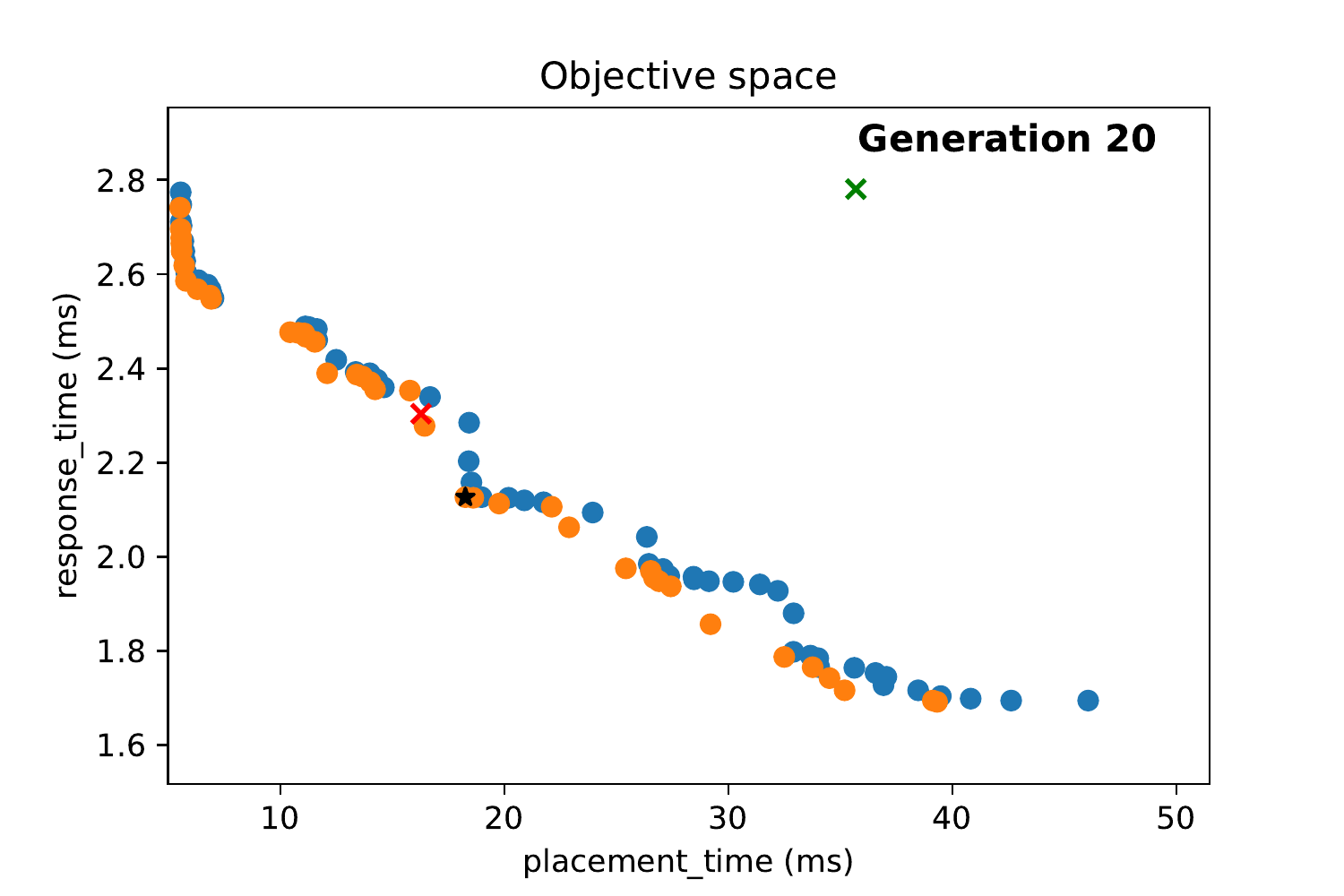}}
	\end{raggedleft}
	\begin{raggedleft} 
	\scalebox{0.45}{\includegraphics[trim=15 5 35 15,clip]{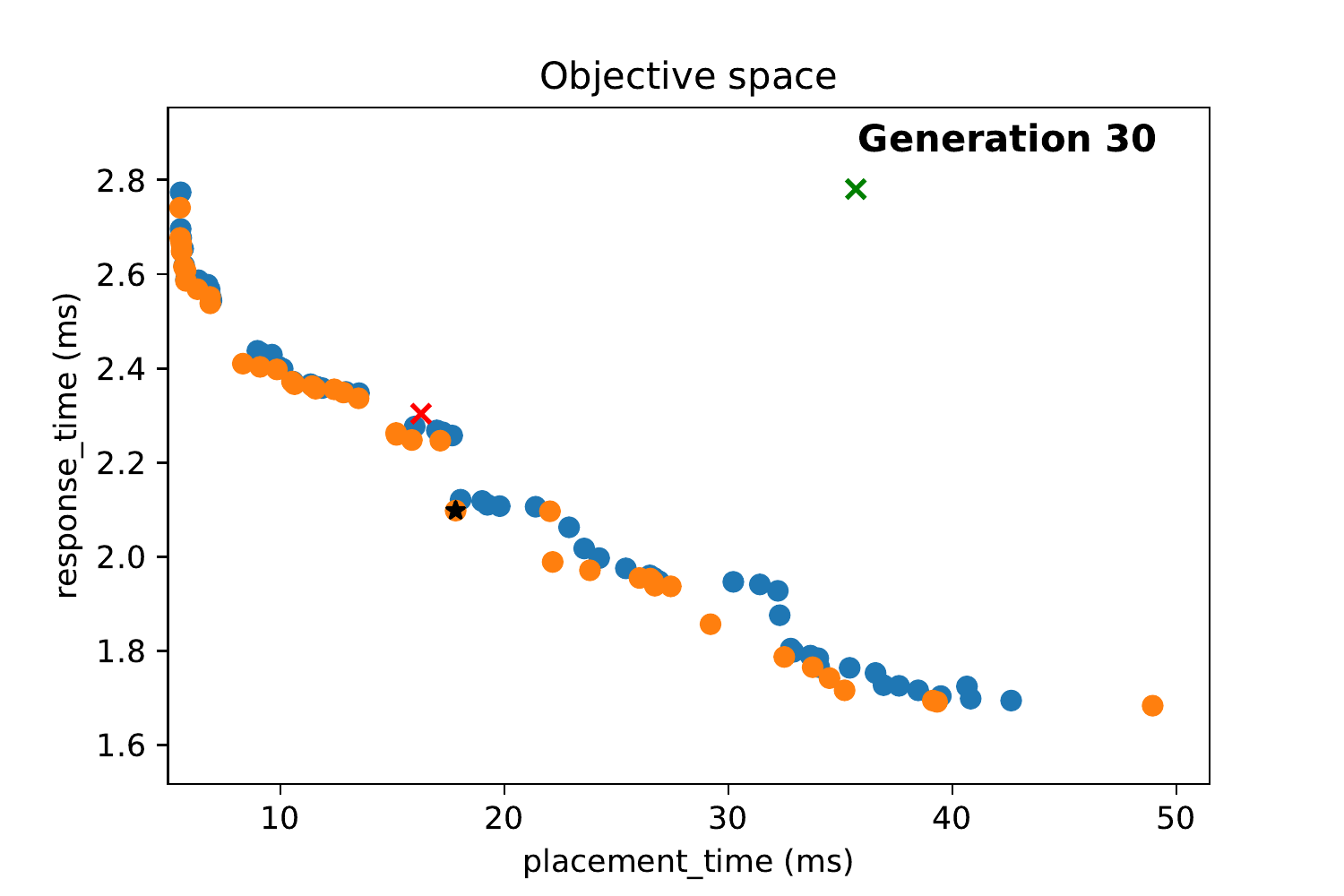}}\hfill
	\scalebox{0.45}{\includegraphics[trim=15 5 35 15,clip]{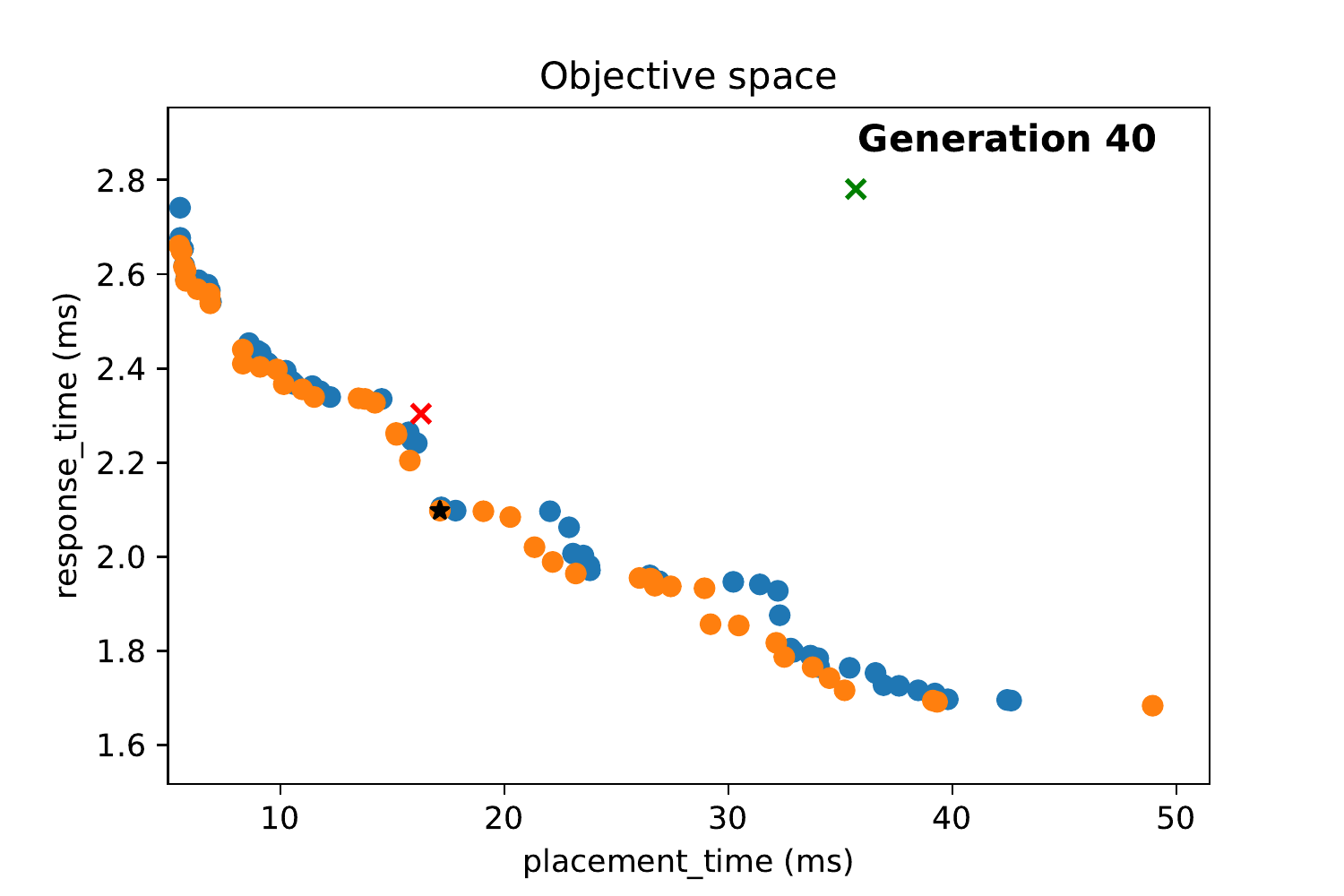}}
	\end{raggedleft}
	\begin{raggedleft} 
	\scalebox{0.45}{\includegraphics[trim=15 5 35 15,clip]{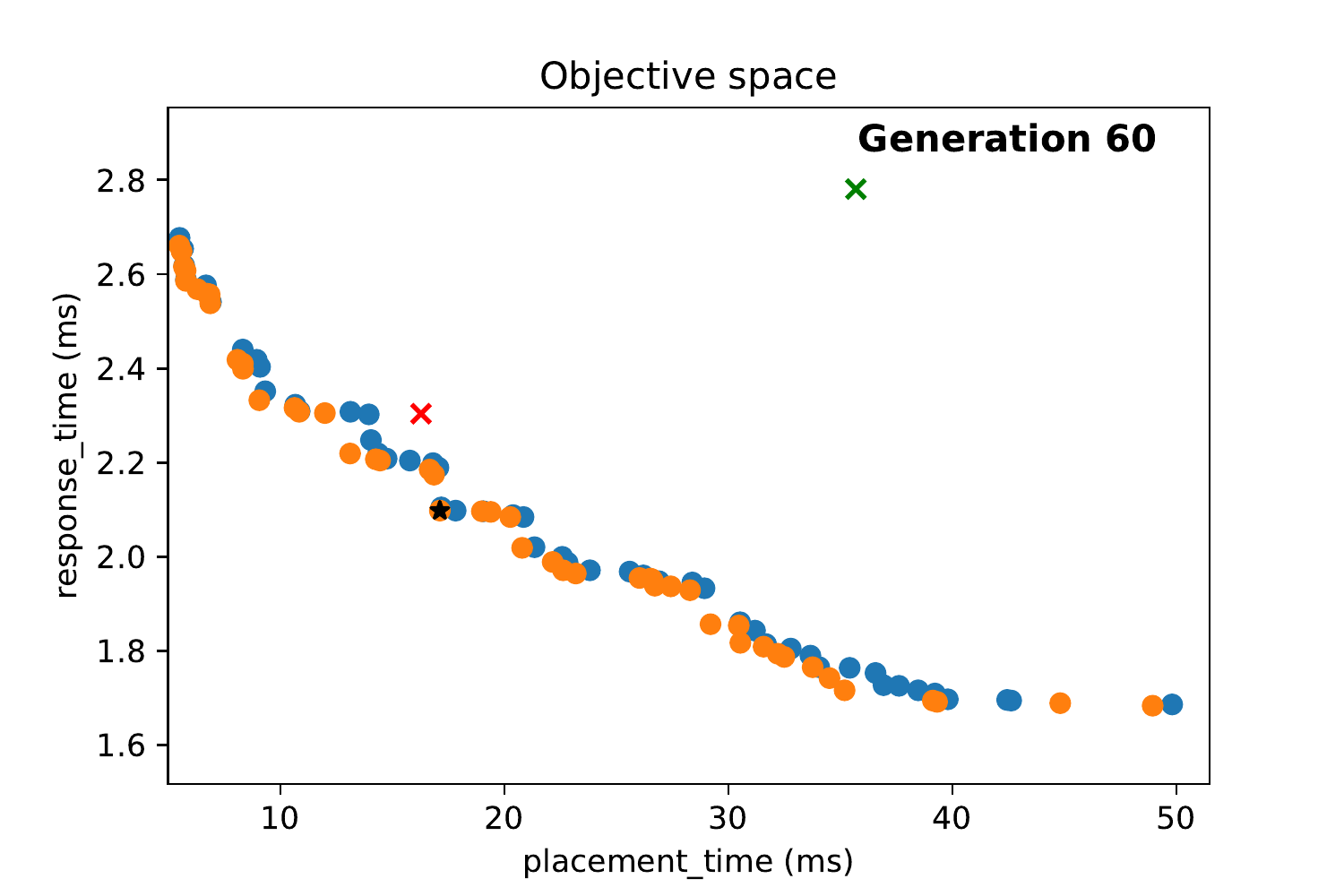}}\hfill
	\scalebox{0.45}{\includegraphics[trim=15 5 35 15,clip]{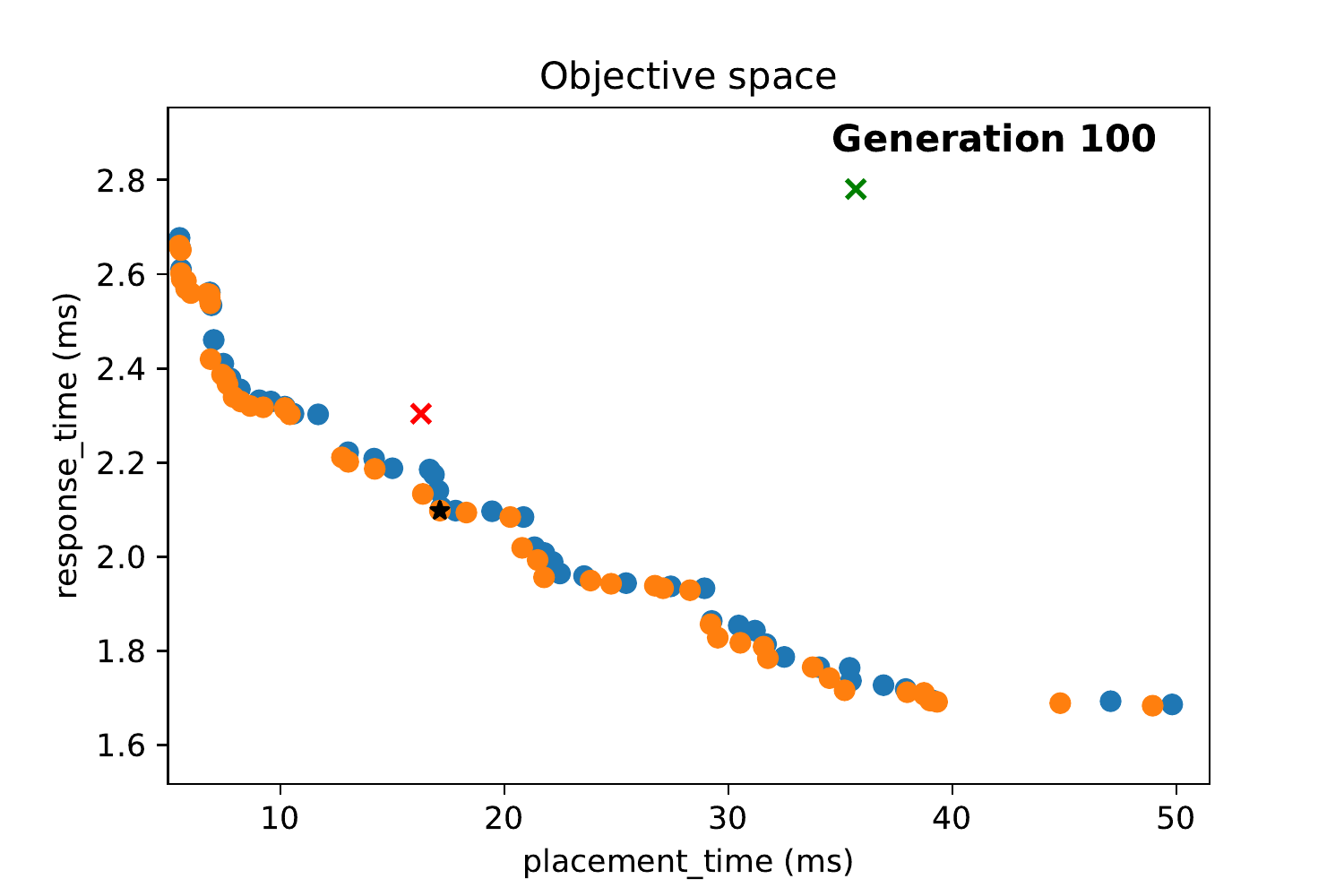}}
	\end{raggedleft}
	\includegraphics{legend.pdf}
	\caption{Objective space evolution of the experiment \textit{200nodes40apps}.}\label{fig_objectivespacesalonggenerations}
\end{figure}

A complete analysis of the evolution of the Pareto front requires to compare it through the GA generations. Figure~\ref{fig_objectivespacesalonggenerations} shows the objective space for different generations along the execution of the GA. Without loss of generality, the plots are limited to 8 generations from the experiment scenario with 200 nodes and 40 apps. The scatter plots for all the generations and all the experiments are available in the source code repository~\footnote{Available on the folder \texttt{results} in \url{https://github.com/acsicuib/GAFogColonies}}.

Figure~\ref{fig_objectivespacesalonggenerations} shows how the number of solutions in the Pareto front is increased (number of orange points) as the generations evolve. Additionally, these solutions increase their homogeneous distribution along the Pareto front as the execution of the optimization progresses. As we explained before, the optimization is better as both the number of solutions in the Pareto front and their homogeneous distribution increase, because it offers a wider range of solutions to the administrator of the system to choose a specific one by applying their criteria. 

It is also observed that the Pareto front evolves by approaching to the origin point of the search space, i.e. by reducing the values of the optimization objectives. Finally, we can observed that the \textit{one-colony} solution (green cross) is not dominated by the Pareto front from the beginning. On contrary, the \textit{fixed-size} solution becomes dominated between generation 20 and 30, as it was expected from the data in Table~\ref{tab_paretocomparative}.

\section{Conclusion}
\label{sect_conclusions}

We have presented a GA to optimize the communication time between users and applications, and to optimize the execution time of the algorithms that perform the placement of the applications into the fog nodes. The optimization is achieved by improving the distribution of the fog devices into fog colonies by taking as reference of this layout the dendrogram of the fog infrastructure.

The experiment results show that defining the fog colony layout with the dendrogram obtained from a hierarchical clustering is suitable to optimize the two performance metrics under study. The results compare the solutions obtained with a multi-criteria implementation of a GA, the NSGA-II, and two control algorithms, defining the colony layout in terms of a fixed size, or considering only one colony that includes all the devices.

By analysing the results, we can conclude that, with an enough number of generations, the Pareto front obtained from the GA dominates the solutions obtained with the two control algorithms. Additionally, the Pareto front included a high number of homogeneously distributed solutions, which allows the system administrator to have a large number of alternatives to choose among.

To sum up, the results of this work confirm: (a) the influence of the fog colony layout on the system performance metrics, (b) the suitability of using a dendrogram to define the fog colony layout, and (c) the suitability of a GA to select  the clusters defined with the dendrogram that optimize two performance metrics.

The future works are mainly focused into improving the optimization process. Firstly, it is required to study if the combination of the GA with other meta-heuristics, creating an hybrid GA, improves the optimization. Secondly, the replacement of the GA with other optimization algorithms needs to be also study, to study which one could obtain the better optimizations.


\section*{Funding}
Funding: Grant PID2021-128071OB-I00 funded by MCIN/AEI/ \\ 10.13039/501100011033 and by the European Union NextGenerationEU/PRTR.

\section*{Data Availability}

The data generated in this study and the source code of the
scripts are public available in the repository \url{https://github.com/acsicuib/GAFogColonies}






\bibliographystyle{elsarticle-num-names} 

\bibliography{bibliography}





\end{document}